\documentclass[a4,structabstract]{aa}  
\usepackage{graphicx}
\usepackage{txfonts}
\usepackage{natbib}
\begin{document}

\title{Spectral signatures of disk eccentricity in young binary systems}
\titlerunning{Spectra of eccentric circumprimary disks}
\subtitle{I. Circumprimary case}
\author{
	Zs. Reg\'aly\inst{1}
    \and
	Zs. S\'andor\inst{2}
	\and
	C. P. Dullemond\inst{2}
	\and
	L. L. Kiss\inst{1,3}
}

\institute{
	Konkoly Observatory of the Hungarian Academy of Sciences, P.O. Box 67, H-1525 Budapest, Hungary\\
	\email{regaly@konkoly.hu}
	\and
	Max Planck Research Group, Max-Planck-Institut f\"ur Astronomie, K\"onigstuhl 17, D-69117 Heidelberg, Germany
	\and
	Sydney Institute for Astronomy, School of Physics A28, University of Sydney, NSw 2006, Australia
}


\abstract
	{Star formation occurs via fragmentation of molecular clouds, which means that the majority of stars born are members of binary systems. There is growing evidence that planets might form in circumprimary disks of medium-separation ($\lesssim\,50\,\mathrm{AU}$) binaries. The tidal forces caused by the secondary generally act to distort the originally circular circumprimary disk to an eccentric one. Since the disk eccentricity might play a major role in planet formation, it is of great importance to understand how it evolves.}
	{We investigate disk eccentricity evolution to reveal its dependence on the physical parameters of the binary system and the protoplanetary disk. To infer the disk eccentricity from high-resolution near-IR spectroscopy, we calculate the fundamental band ($4.7\,\mathrm{\mu m}$) emission lines of the CO molecule emerging from the atmosphere of the eccentric disk.}
	{We model circumprimary disk evolution under the gravitational perturbation of the orbiting secondary using a 2D grid-based hydrodynamical code, assuming $\alpha$-type viscosity. The hydrodynamical results are combined with our semianalytical spectral code to calculate the CO molecular line profiles. Our thermal disk model is based on the double-layer disk model approximation. We assume LTE and canonical dust and gas properties for the circumprimary disk.}
	{We find that the orbital velocity distribution of the gas parcels differs significantly from the circular Keplerian fashion. The line profiles are double-peaked and asymmetric in shape. The magnitude of asymmetry is insensitive to the binary mass ratio, the magnitude of viscosity ($\alpha$), and the disk mass. In contrast, the disk eccentricity, thus the magnitude of the line profile asymmetry, is influenced significantly by the binary eccentricity and the disk geometrical thickness.}
	{We demonstrate that the disk eccentricity profile in the planet-forming region can be determined by fitting the high-resolution CO line profile asymmetry using a simple 2D spectral model that accounts for the velocity distortions caused by the disk eccentricity. Thus, with our novel approach the disk eccentricity can be inferred from high-resolution near-IR spectroscopy data acquired prior to the era of high angular resolution optical (ELT) or radio (ALMA, E-VLA) direct-imaging. By determining the disk eccentricity in medium-separation young binaries, we might be able to constrain the planet formation theories.}


\maketitle

\section{Introduction}

Star formation occurs via fragmentation of molecular clouds causing about 60\% of stars to be born as a member of a binary system \citep{DuquennoyMayor1991}. 
Large initial specific angular momentum results in circumbinary disk formation around the protobinary, while circumstellar disks (a circumprimary and a circumsecondary) are formed around the protostars for lower initial specific angular momentum \citep{BateBonnell1997}. \citet{BonavitaDesidera2007} demonstrated that the overall frequency of giant planets in binaries and single stars does not statistically differ among planets discovered by radial velocity surveys. On the basis of a comprehensive survey for companions of 454 nearby Sun-like stars, \citet{Raghavanetal2010} revealed that both single and multiple stars are equally likely to harbor planets. 

Most of the planet-bearing binaries have large separations where the planet formation might be unaltered by the companion's gravitational perturbation. However, stellar multiplicity might play a major role in planet formation in medium-separation ($\lesssim 50\,\mathrm{AU}$) binary systems. Based on Doppler surveys, \citet{EggenbergerUdry2010} showed that about 17\% of circumstellar exoplanets are associated with binaries. Among this, five circumstellar exoplanets are known to date in $\lesssim50\,\mathrm{AU}$ separation binary system \citep{Quelozetal2000,Hatzesetal2003,Zuckeretal2004, Lagrangeetal2006,Chauvinetal2006,Correiaetal2008}. Two exceptional cases are also known, HW Virginis \citep{Leeetal2009} and CM Draconis \citep{Deegetal2008}, in which planets have been detected in circumbinary orbits. Thus, planet formation theories, such as core-accretion \citep{BodenheimerPollack1986,Pollacketal1996} or gravitational instability \citep{Boss2001}, must be able to explain the formation of planets in both circumprimary and circumbinary disk environments. 

The circumprimary disk is tidally truncated at $0.35-0.5$ times the binary separation, depending on the binary mass ratio, binary eccentricity, and magnitude of the disk viscosity \citep{ArtymowiczLubow1994}. Owing to the angular momentum transfer between the disk and the companion \citep{PapaloizouPringle1977}, the disk is truncated, resulting in a greatly reduced disk lifetime being available for planets to form in $\lesssim50\,\mathrm{AU}$ binaries \citep{Ciezaetal2009} that is a severe problem for core-accretion scenario. In contrast, the formation of gas giant planets by the relatively rapid-action of the gravitational instability might be induced by the secondary-generated shock waves if the gas cooling time is short. Nevertheless, the disk viscosity can heat the disk sufficiently to suppress the formation of clumps \citep{Nelson2000}, but with small viscosity, gravitationally unstable clumps can still form \citep{Boss2006}.

The disk gas feels the companion's periodic perturbation leading to strong interaction at the location of the Lindblad resonances. Waves launched at Lindblad resonances carry energy and angular momentum from the binary. The disk experiences changes in its angular momentum where the waves dampen, resulting in the development of an eccentric disk \citep{Lubow1991}. Several mechanism that may cause wave damping have been proposed, e.g. shocks that could be effective in colder disks, or turbulent disk viscosity acting as a dissipation source and radiative damping. Both SPH \citep{ArtymowiczLubow1994} and grid-based \citep{Kleyetal2008} hydrodynamical simulations have confirmed the eccentricity development in binaries assuming spatially constant viscosity for the gas as a source of the wave damping. Since the orbit of bodies (dust particles or pebbles) is perturbed not only by the periodic gravitational potential of the binary but the gas drag as well, the development of disk eccentricity might influence the core-accretion processes.

The maximum size of the building blocks of planetesimals is affected by the impact velocity of sub-micron-sized grains in the dust coagulation process. According to Zsom et al. (2010), the aggregate sizes are lower in eccentric disks than in axisymmetric disk environments owing to the increase in the relative velocity between the dust particles. We note, however, that an investigation of the SED slopes of medium-separation T\,Tauri binaries by \citet{Pascuccietal2008} showed that the extent of dust processing in the disk surface layer and the degree of dust settling in binary disks do not differ significantly from those in disks around single stars. 

The planetesimal accretion phase, leading to between km-sized planetesimals and several 100\,km-sized planetary embryos, should proceed in an environment where the mutual encounter velocities of planetesimals are on the order of planetesimal surface escape velocities. In this environment, the planetary embryos could grow quickly by runaway growth mode \citep{WetherillStewart1989}. Since the runaway growth mode is sensitive to the encounter velocities, for increased encounter velocities, e.g., due to the stirring up of the planetesimal swarm, runaway growth might be stopped. \citet{Thebaultetal2006} showed that the impact velocity of different-sized planetesimals tends to increase owing to the interaction between the companion and the gaseous friction in binaries with separations of $10\leq a_\mathrm{bin}\leq50\,\mathrm{AU}$. \citet{Paardekooperetal2008} found that the planetesimal encounter velocities with different sizes could be larger by an order of magnitude in eccentric disks than in the axisymmetric case. Consequently, planetesimal accretion might be inhibited in highly eccentric disks. 

In accretion disks, the presence of the double-peaked emission lines are the natural consequence of the gas parcels moving in Keplerian orbits around the host star. \citet{Huang1972} and \citet{Smak1981} presented this in connection with the emission lines of Be and cataclysmic variable stars. \citet{HorneMarsh1986} investigated the emerging line profiles in accretion disks. Since the Keplerian angular velocity of gas parcels is highly supersonic in accretion disks, the Doppler shift of the line emitted by individual gas parcels exceeds the local line-profile width. Summing up the line profiles emitted by individual rings of gas parcels, and taking into account the radial dependence of the line surface brightness, the result is the well-known double-peaked broad symmetric line shapes \citep{HorneMarsh1986}. Azimuthal asymmetries in disk surface brightness (e.g. density perturbations in optically thin lines, or supersonic anisotropic turbulence in saturated lines), will break the line profile symmetry \citep{Horne1995}. Gas parcels orbiting non-circularly (i.e., in elliptic orbit) might also produce the asymmetric line profiles presented by \citet{Foulkesetal2004} for cataclysmic variables, and Reg\'aly et al., (2010) for protoplanetary disks.

In this paper, we investigate the eccentricity evolution of a circumprimary disk in a young binary system. We perform an extensive parameter study to reveal the dependence of the disk eccentricity on several parameters, such as binary and disk geometry, and gas viscosity. The hydrodynamical simulations were done in 2D by a grid-based parallel hydrodynamic code FARGO \citep{Masset2000}. We model the circumprimary disk evolution under the gravitational perturbation of the orbiting secondary assuming $\alpha$-type viscosity \citep{ShakuraSunyaev1973}. We calculate the fundamental band ($4.7\,\mathrm{\mu m}$) ro-vibrational emission lines of the molecule $\mathrm{^{12}C^{16}O}$ emerging from the disk atmosphere, providing a tool to determine the disk eccentricity from high-resolution near-IR spectroscopy by means of line profile distortions. Our thermal disk model is based on the double-layer disk model of \citet{ChiangGoldreich1997}. Since the velocity distribution of the gas parcels show supersonic deviations from the circular Keplerian one, owing to the eccentric disk state, asymmetric molecular line profiles emerge from the optically thin disk atmosphere. 

The paper is structured as follows. In the next section, we present our hydrodynamical simulation modeling of the evolution  in general of disk eccentricity. The calculation of fundamental-band CO ro-vibrational emission lines emerging from an eccentric circumprimary disk are presented in Sect. 3. In Sect. 4, we present an extensive parameter study to investigate the evolution of the disk eccentricity for a wide range of binary and disk parameters. Section 5 deals with the comparison of our results to other recent simulations, and the observability of eccentric signatures. The paper closes with conclusions.

\section{Hydrodynamic disk model}

In our simulations, we use a locally isothermal version of FARGO, a publicly available parallel 2D hydrodynamical code \citep{Masset2000}. We apply $\alpha$-type disk viscosity \citep{ShakuraSunyaev1973}, assuming $\alpha=0.02$ in our model. We adopt dimensionless units for which the unit of length is the orbital separation of the binary ($a_\mathrm{bin}$), and the unit of mass is the mass of central star. The unit of time $t_0$ is obtained from the orbital period of the binary, thus $t_0=1/2\pi$, setting the gravitational constant $G$ to unity. 

The computational domain is covered by 270 radial and 500 azimuthal grid cells, and the origin of the grid is on the primary star. The radial spacing is logarithmic, while the azimuthal spacing is equidistant. The disk's inner and outer boundary is at $0.01a_\mathrm{bin}$ and $1.5a_\mathrm{bin}$, respectively. The initial orbital separation of the binary is $a_\mathrm{bin}=1$, thus the computational domain contains the orbit and the Roche lobe of the system. Open boundary conditions are assumed at the inner and outer boundaries, i.e the disk material is allowed to flow out from the computational domain, but no inflow is allowed. The secondary is allowed to accrete the material flowing through its Hill sphere with the same rate as described in \citet{Kley1999}. The secondary feels the gravitational pull of the disk, thus its orbital elements can change.

We use a rotating frame that co-rotates with the binary. The binary orbit is initially circular ($e_\mathrm{bin}=0$).  The initial surface density profile follows a power-law distribution $\Sigma(R)=\Sigma_0R^{-0.5}$. We use the canonical constant aspect ratio $h=0.05$ disk approximation. The disk self-gravity is neglected because the Toomre parameter $Q(R)=hM_\mathrm{*}/\pi R^2\Sigma(R)\gg1$ for our model \citep{Toomre1964}.

To model an existing binary system V807\,Tau with a sole circumprimary disk \citep{HartiganKenyon2003}, the secondary-to-primary mass ratio is set to 0.3. The binary eccentricity is neglected in this particular calculation. The distance unit is taken to be $40\,\mathrm{AU}$ accordingly to the measured separation \citep{Pascuccietal2008}. The initial disk mass is set to $2\times 10^{-3}\,M_{\odot}$ by the choice of $\Sigma=17.2\,\mathrm{g/cm^2}$ at 1\,AU. To prevent numerical instabilities at low density regions, an artificial density floor is applied: whenever the disk surface density is below a certain limit, it is reset to that limit, which is $10^{-12}$ in dimensionless units.

Initially, we placed the secondary on a circular orbit into the unperturbed disk, in which the surface density had been artificially damped. With this damping, the secondary orbits in a practically gas-free environment. We use a Gaussian damping, i.e., the density is damped as
\begin{equation}
	\label{eq:truncation}
	\Sigma(R,\phi)_\mathrm{trunc}=\Sigma(R,\phi)\left( \Sigma_\mathrm{s}+(1-\Sigma_\mathrm{s})\exp\left[-\left(\frac{R-R_\mathrm{tr}}{\sigma_\mathrm{tr}}\right)^2\right]\right),
\end{equation}
where $\Sigma_\mathrm{s}$ is the surface density at the outer edge of the damped region, $R_\mathrm{tr}$ is the radius where the density damping begins, and $\sigma_\mathrm{tr}$ is the radial extent of the damped region. The damping parameters are $R_\mathrm{tr}=0.6$, $\sigma_\mathrm{tr}=0.2$, and $\Sigma_\mathrm{s}=10^{-8}$.

\subsection{Formation of eccentric disk}

\begin{figure}
	\centering
	\includegraphics[width=\columnwidth]{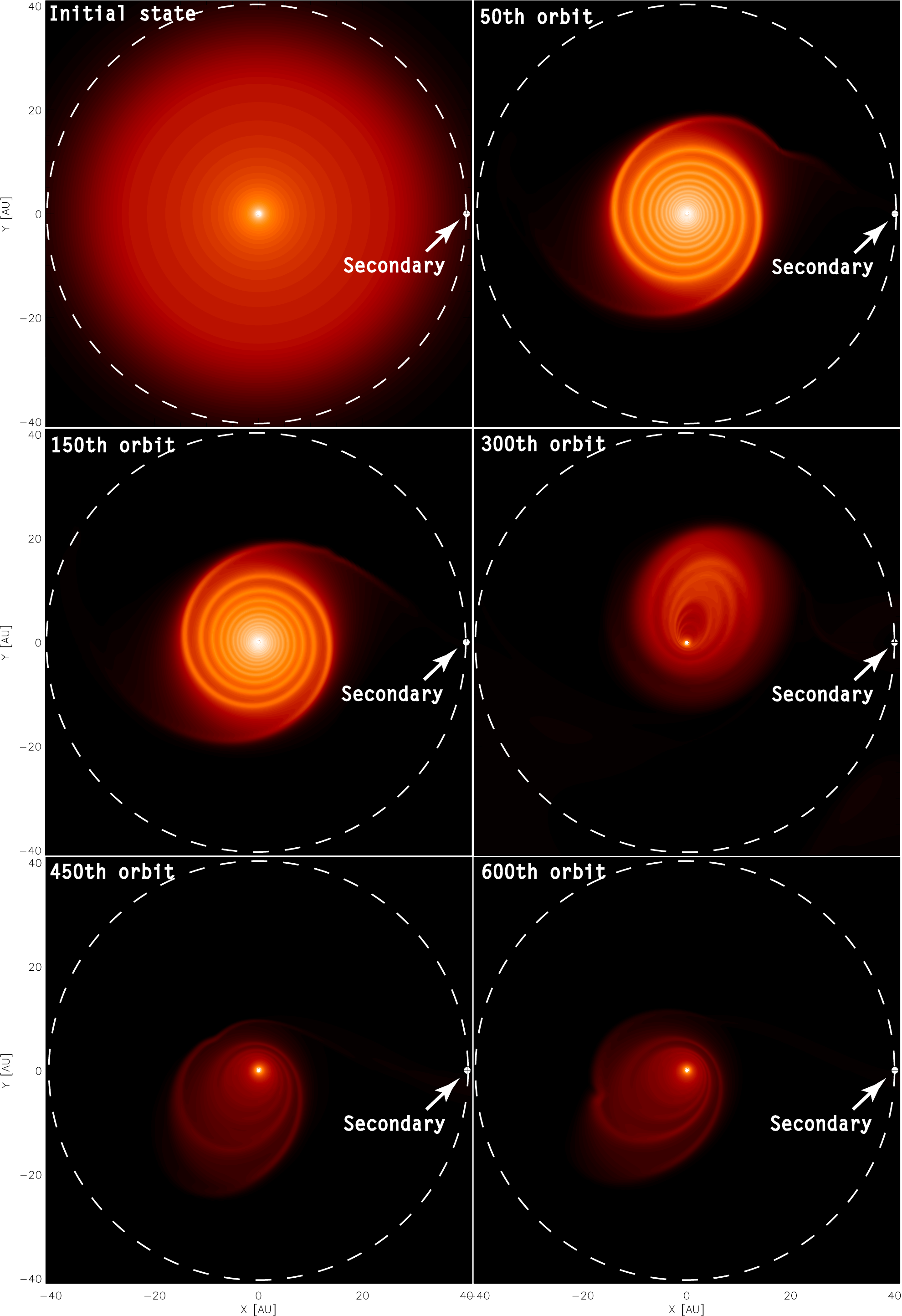}
	\caption{Evolution of the disk surface density shown in six snapshots taken at 0, 50, 150, 300, 450, and 600th orbits of the binary. The secondary's counter-clockwise orbit is shown with a white dashed circle.}
	\label{fig:densevol}
\end{figure}

\begin{figure*}
	\centering
	\includegraphics[width=\textwidth]{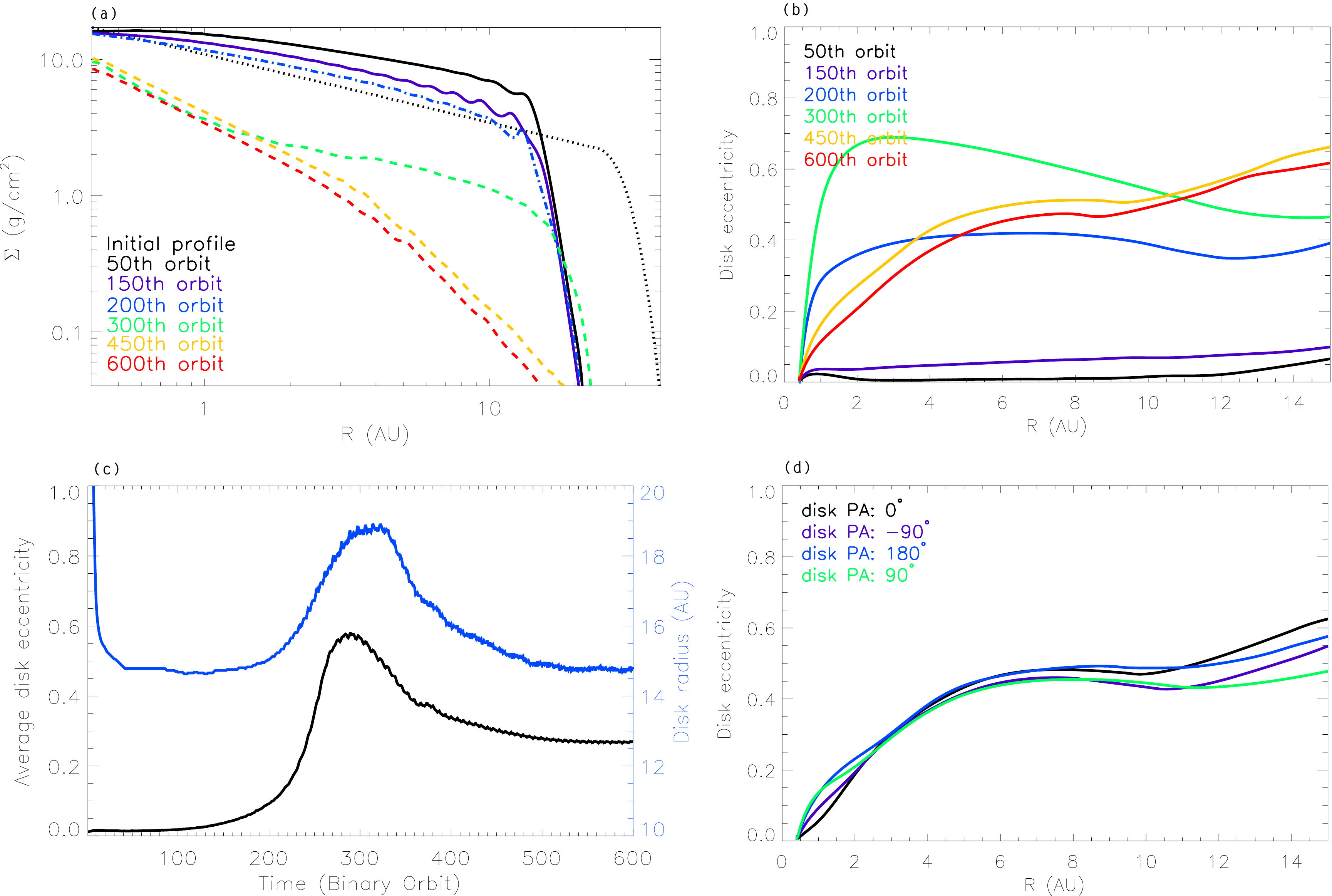}
	\caption{\emph{Panel a}: evolution of the azimuthally averaged density profile shown in six epochs calculated at 0, 50, 150, 300, 450, and 600th binary orbits. The initial density profile is shown with black dotted curve. \emph{Panel b}: evolution of the azimuthally averaged eccentricity shown in five curves calculated at 50, 150, 300, 450, and 600th binary orbits. Only the 0-20\,AU region is shown in the plot, as the disk is significantly truncated beyond 20\,AU. \emph{Panel c}: evolution of the disk radius (with blue), and the averaged disk eccentricity (with black). The curves are smoothed to remove ``noise'' (see text for explanation). \emph{Panel d}: variation in the disk eccentricity during a full disk precession period. The disk position angle (PA) is measured relative to the position of the orbiting secondary.}
	\label{fig:profile_evol}
\end{figure*}

We present our results on the disk evolution through a couple of hundred orbits of the binary. The evolution of the 2D surface density is shown in Fig.\,\ref{fig:densevol}, and the azimuthally averaged density profile in Fig.\,\ref{fig:profile_evol} (panel a).

After the first couple of binary orbits, double spiral waves appear. In general, the disk remains approximately axisymmetric until the $\sim 150$th binary orbit (Fig.\,\ref{fig:densevol}). Comparing the initial and the evolved density profiles in the 50th and 150th orbits, the profile becomes less steep, while the density at the disk inner edge increases as the disk material is piled up by the secondary (Fig.\,\ref{fig:profile_evol}, panel a). The density profile is steepened during the subsequent orbits because of the angular momentum removal by the secondary.

The disk eccentricity is calculated at each individual grid cells using the radial and azimuthal velocity components applying the equations presented in \citet{Regalyetal2010}. To obtain the radial disk-eccentricity profiles, the eccentricity distribution is azimuthally averaged within the disk radius, which is defined as the radius containing 90\% of the mass. From the 150th orbit, the disk eccentricity rapidly increases (see the evolution of disk eccentricity in Fig.\,\ref{fig:profile_evol} panel b). It is expedient to radially average the disk-eccentricity profiles at each secondary orbit, in which case we can assign a single value to each eccentric state of the disk. Since the disk radius and eccentricity profile (Fig.\,\ref{fig:profile_evol}, panel d) slightly change as the disk and the secondary mutual positions vary, their values as a function of time are ``noisy''. To remove this ``noise'', the radius and eccentricity evolution curves calculated at each secondary orbit are smoothed (Fig.\,\ref{fig:profile_evol}, panel c).

At the beginning of the simulation, the disk is truncated by the secondary to $\sim15\,\mathrm{AU}$. From the 200th binary orbit, the disk begins to expand slightly, and the disk eccentricity starts to grow. The disk eccentricity reaches its overall maximum during the $\sim 300$th binary orbit, and begins to decline during subsequent orbits. The disk reaches a quasi-steady state by the $\sim450$th binary orbit, in which neither the eccentricity nor the disk radius evolves further. By this time, the disk radius is stabilized at $\sim15\,\mathrm{AU}$. Noteworthy is the correlation of the disk eccentricity and radius evolution.

Regarding the orbital parameters of the secondary, we found that they are not changed significantly. Until the 600th orbit, the secondary migrates slightly outward to 0.3\% of its original distance because it gains angular momentum from the disk, while the binary eccentricity remains close to zero. 

\begin{figure}
	\centering
	\includegraphics[width=\columnwidth]{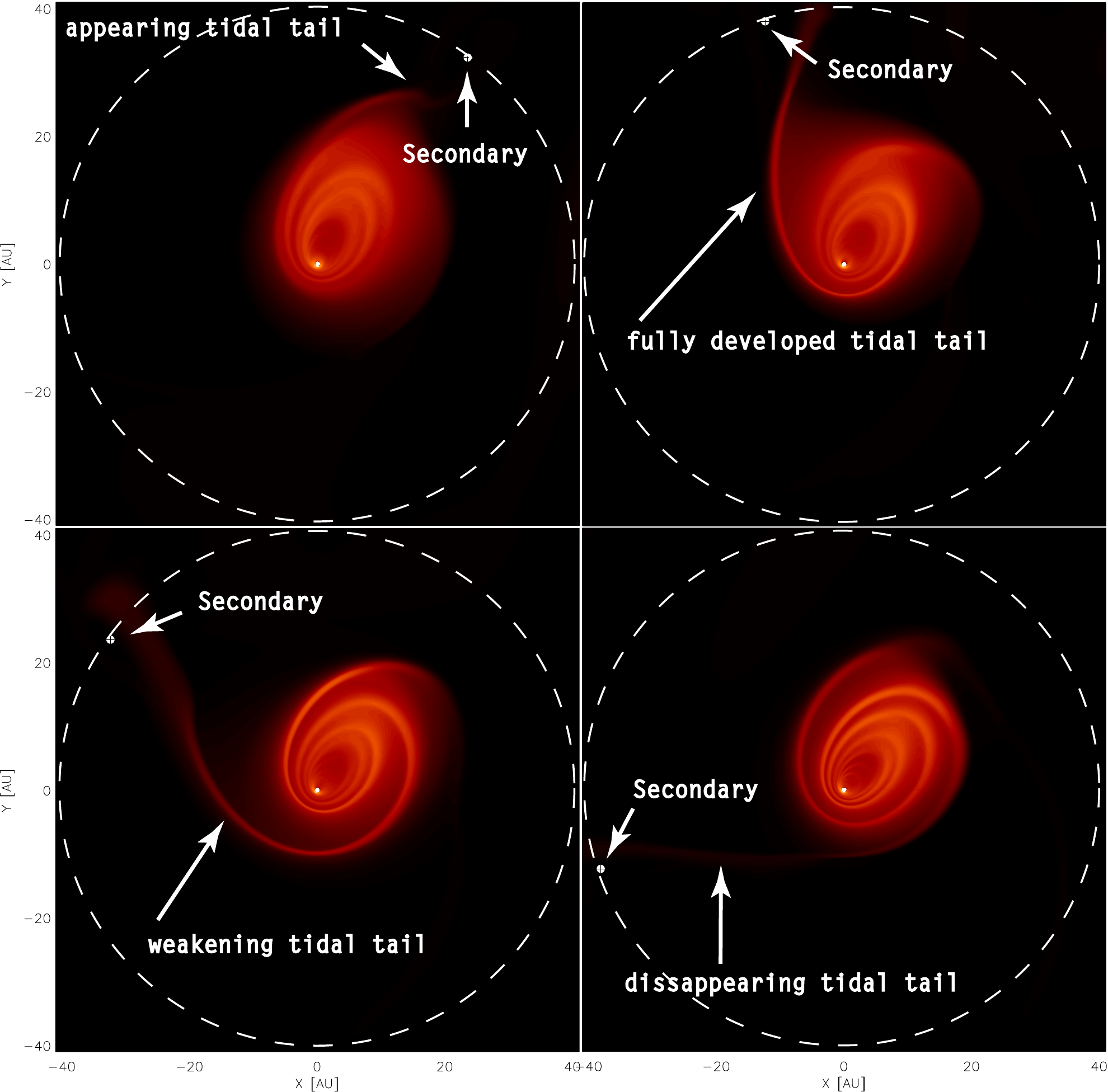}
	\caption{Development of the tidal tail appeared after the secondary passes closest to the disk apastron edge during the 300th orbit. The secondary's counter-clockwise orbit is shown with a white dashed circle. The density enhancement can reach two orders of magnitude in the tidal tail.}
	\label{fig:sm_flow}
\end{figure}

When the secondary passes close to the outer edge of the elliptic disk, a tidal tail develops through which the disk material increases the mass of the secondary, but by a negligible amount. The evolution of a tidal tail during the 300th orbit is shown in Fig.\,\ref{fig:sm_flow}. The tidal tail persists for only a quarter of an orbit, and reappears in each binary revolution. 

\begin{figure}
	\centering
	\includegraphics[width=\columnwidth]{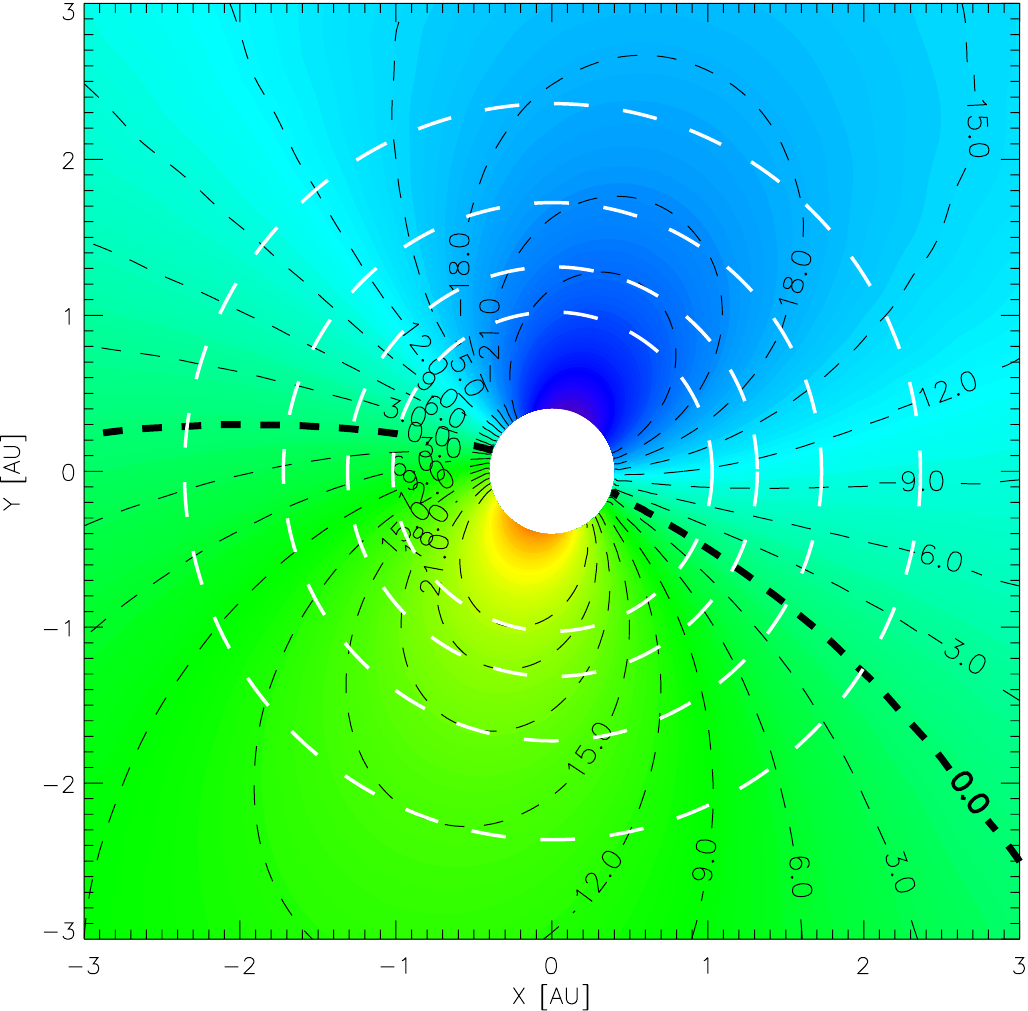}
	\caption{Line-of-sight velocity component of the gas parcels in the 600th orbit, assuming a $i=60^\circ$ disk inclination. The unit of contour lines is $km\,s^{-1}$. The reddish and bluish colors represent the receding and approaching part of the disk, respectively. Gas parcels at the same distance to the primary (\emph{white dashed lines}) on the approaching side have significantly larger apparent velocities than those on the receding side.}
	\label{fig:azimvel}
\end{figure}

When the disk becomes elliptic, the velocity field departs significantly from the circular Keplerian one. The disk is non-axysimmetric, not only in density, but in velocity field too. The line-of-sight velocity component of the gas parcels calculated during the 600th binary orbit is shown in Fig.\,\ref{fig:azimvel}, assuming that the disk is seen to be inclined by $60^\circ$. Gas parcels that are at the same distance to the primary star (shown with white dotted circles), hence at the same temperature, have different azimuthal velocities. Since the velocity asymmetry at a given radius can be larger than the local sound speed ($2-3\,\mathrm{km\,s^{-1}}$) for $R\leq3\,\mathrm{AU}$, the Doppler shift of the lines  emitted by a gas parcel exceeds the local line width. Consequently, the shape of emission lines emerging from the eccentric disk might be distorted \citep{HorneMarsh1986}. Therefore, we expect the molecular line profiles formed in an eccentric circumprimary disk of a young binary system to be asymmetric (see, e.g., \citet{Statler2001}).

The elliptic disk precesses retrogradely with respect to the secondary's orbit, as if it were a rigid body. The precession period is $\sim 6.6$ orbital periods of the binary. As a consequence, the line profiles are expected to exhibit long-period variations. 

The azimuthally averaged disk eccentricities calculated for four different disk position angles (PA) measured with respect to the secondary are shown in Fig.\,\ref{fig:profile_evol}, panel d. The eccentricity of the inner disk ($R\leq3\,\mathrm{AU}$) shows periodic variation. The disk eccentricity reaches its maximum, when the disk is aligned with the binary at $\mathrm{PA}=180^\circ$, i.e., when the disk periastron edge is the closest to the secondary. When the disk apastron edge is closest to the binary, i.e., for $\mathrm{PA}=0^\circ$, the disk eccentricity reaches its overall minimum. Between these states, when the disk semimajor axis is perpendicular to the binary axis, the inner-disk average eccentricity has an intermediate value. As a consequence, the line profiles also are expected to display small periodic variations on the timescale of the binary orbit.

\section{Fundamental band ro-vibrational CO line profiles}

The spectral lines emerging from an axisymmetric disk with disk parcels on circular Keplerian orbits, have symmetric double-peaked shapes \citep{HorneMarsh1986}. Analyzing the hydrodynamical results, it is reasonable to expect that the optically thin atmosphere of an eccentric disk produces non-symmetric molecular lines as the disk parcels are moving on non-circular orbits. We note that several single T\,Tauri stars display centrally peaked symmetric CO line profiles. This might be attributed to an extended CO excitation by stellar UV photons (see e.g., \citet{Brittainetal2007}), in which case the line profiles are also expected to be asymmetric. In the following, we do not take into account the stellar UV flux, thus our investigation concentrates on systems that would produce double-peaked profiles.

To calculate the strengths of the CO ro-vibrational emission lines, we combine the hydrodynamic results with our recently developed semi-analytical line spectral model (Reg\'aly et al., 2010). In our approach, the thermal disk model is based on the double-layer disk approximation of \citet{ChiangGoldreich1997}, in which the CO emission lines are formed in the superheated optically thin disk atmosphere above the cooler optically thick disk interior. The disk inner edge, where the simple double-layer assumption cannot be applied, is assumed to be a perfectly vertical wall. According to the stellar evolutionary tracks \citep{Siessetal2000}, the luminosity of a $0.3\,M_\odot$ pre-main-sequence star is negligible compared to that of an $1\,M_{\odot}$ primary ($L_{0.3M_{\odot}}/L_{1M_{\odot}}\simeq0.2$), hence we neglect the irradiation from the secondary. We therefore assumed that the disk is heated solely by the stellar irradiation of the primary. We note that the accretion rate measured in our hydrodynamical simulations is below $2.5\times10^{-11}\,M_{\odot}\mathrm{yr^{-1}}$, thus viscous heating is negligible compared to the stellar irradiation. As a consequence, the disk atmosphere is hotter than the disk interior, where a superheated atmosphere is formed producing emission molecular spectra. We assume that the primary is similar to the one in the system V807\,Tau. On the basis of the published spectral type of V807\,Tau, i.e., K7 \citep{Hartiganetal1994} and assuming that the primary is $\sim2.5\,\mathrm{Myr}$ old, its radius and effective surface temperature are taken to be $R_*=1.83\,R_{\odot}$, and $T_*=4266\,\mathrm{K}$, respectively, using a publicly available pre-main-sequence database  \citep{Siessetal2000}. It is assumed that the dust consists of pure silicates with a $0.1\,\mathrm{\mu m}$ grain size \citep{DrainLee1984}. The mass absorption coefficient of the dust is taken to be $\kappa^\mathrm{dust}_\mathrm{V}=2320\,\mathrm{cm^2/g}$ at visual wavelengths and $\kappa^\mathrm{dust}_\mathrm{4.7\mu m}=200\,\mathrm{cm^2/g}$ at $4.7\mathrm{\mu m}$. The dust-to-gas and CO-to-gas mass ratios are assumed to be constant throughout the disk and being assumed to be $X_\mathrm{d}=10^{-2}$ and $X_\mathrm{g}=4\times10^{-4}$, respectively.

Taking into account our dust opacity assumptions, the disk is optically thick within 3\,AU where the CO ro-vibrational fundamental band is excited, regardless of the significant depletion that occurs by the time the disk eccentricity has developed (see Fig.\,\ref{fig:profile_evol}, panel a). The continuum optical depth normal to the disk midplane at 4.7$\mathrm{\mu m}$ is $\tau_\mathrm{4.7\mu m}(R)=\Sigma(R)X_\mathrm{d}\kappa^\mathrm{dust}_\mathrm{4.7 \mu m}$, which drops below 1 (for one side of the disk) if the disk surface density is $\Sigma_\mathrm{d}\leq 0.5\mathrm{g/cm^2}$. In our simulation, the surface density  is $\Sigma\geq1\mathrm{g/cm^2}$ within 3\,AU (see Fig.\,\ref{fig:profile_evol}, panel c), therefore the disk interior is optically thick at both 4.7\,$\mu$m and optical wavelengths. However, the density drops below this critical value beyond 3\,AU (see Fig.\,\ref{fig:profile_evol} panel a), thus the disk is no longer optically thick in the outer regions.\footnote{Note, however, that the density can be arbitrarily scaled up resulting in an optically thick disk, as the hydrodynamical simulations are independent of the assumed disk mass as long as the change in binary orbital elements are small. More details are given in the Discussion.} Nevertheless, beyond 2-3\,AU the CO ro-vibrational fundamental band is not excited by the primary irradiation, and beyond 5\,AU the disk interior has no contribution to the 4.7$\mu$m continuum.

\begin{figure*}
	\centering
	\includegraphics[width=18cm]{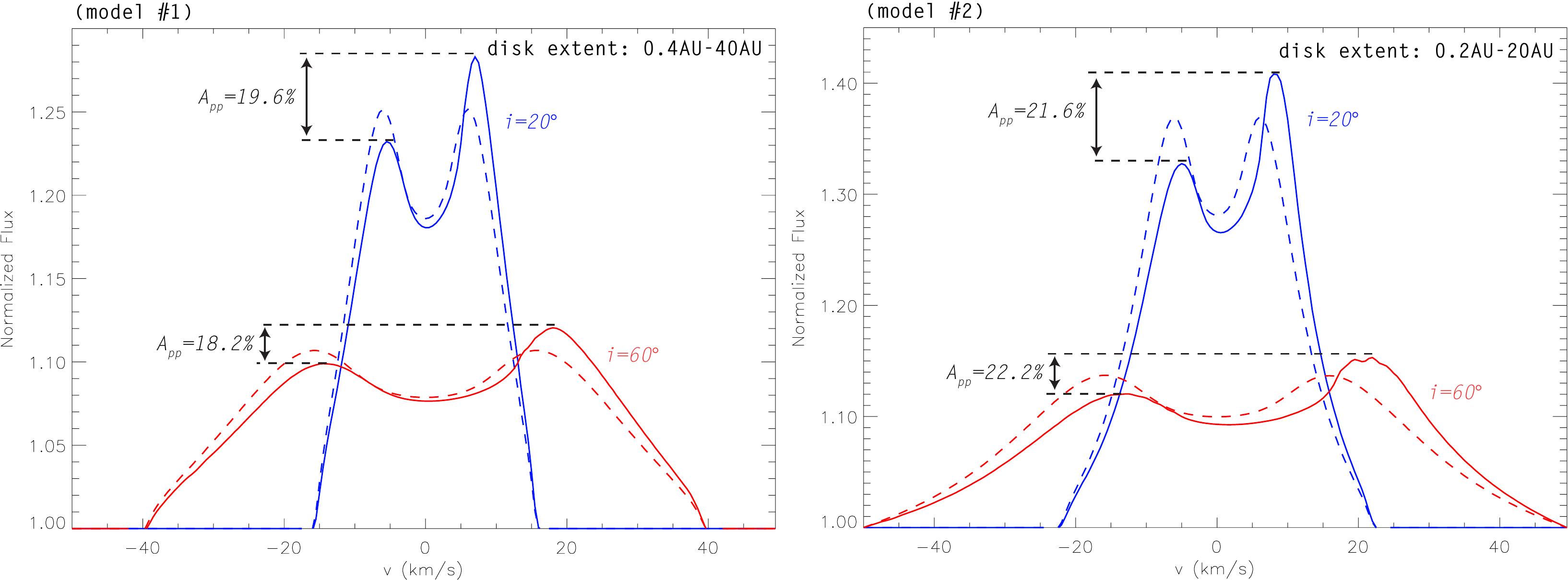}
	\caption{$V=1\rightarrow 0$ P10 fundamental band CO ro-vibrational line profiles for circular Keplerian (\emph{dashed lines}) and eccentric disk (\emph{solid lines}) in the 600th binary orbit. The disk extensions were 0.4\,AU-40\,AU and 0.2\,AU-20\,AU for panel \emph{left} and \emph{right}, respectively. The line profiles for inclinations $i=60^{\circ}$ and  $i=20^{\circ}$ are shown with blue and red colors, respectively. The magnitude of asymmetry is larger for higher inclination angles. Although the line fluxes normalized to the continuum are clearly different, the magnitude of asymmetry remains the same for model \#1 and \#2.}
	\label{fig:V1-0P10} 
\end{figure*}

\subsection{Asymmetric line profiles}

In advance, here we give a short summary of the physical background responsible for the shaping of molecular line profiles emerging in dynamically distorted disk atmospheres. The local line profile emerging from the disk atmosphere from a given point $R,\phi$ in cylindrical coordinates depends on the local Doppler shift. The fundamental line center $\nu_0$ shifts because of the apparent motion of the gas parcels along the line-of-sight. The shift is significant compared to the line width if the apparent velocity of the gas parcels exceeds the local sound speed.\footnote{We neglect the effect of turbulent line broadening, thus the intrinsic line width is determined by the thermal broadening alone.} To calculate the Doppler shifts for circularly Keplerian disks and an eccentric disk, we used Eq.\,(E.8) and Eq.\,(4) of \citet{Regalyetal2010}, respectively. In the latter case the velocity components of the gas parcels ($u_{\mathrm{R}}(R,\phi)$ and $u_{\mathrm{\phi}}(R,\phi)$) were provided by the hydrodynamical simulations.

\begin{table}
\begin{minipage}[t]{\columnwidth}
\caption{CO line profile asymmetries for model \#1 and \#2}
\label{table:lp-asymmetry}
\centering
\renewcommand{\footnoterule}{}
\begin{tabular}{c c c c c}
\hline\hline
Model & $i$ & $I_b^\mathrm{a}$ & $I_{r}^\mathrm{b}$ & $A_\mathrm{pp}^\mathrm{c}$ \\
\hline
\#1  & $20^{\circ}$ & 1.23 & 1.28 & 19.6\%   \\
& $60^{\circ}$ & 1.10 & 1.12 & 18.2\% \\
\hline                         
\#2  & $20^{\circ}$ & 1.33 & 1.41 & 21.6\% \\
& $60^{\circ}$ & 1.12 & 1.15 & 22.2\%   \\
\hline
\end{tabular}
\begin{list}{}{}
\item[$^{\mathrm{a}}$] Normalized intensity maximum at the red peak
\item[$^{\mathrm{b}}$] Normalized intensity maximum at the blue peak
\item[$^{\mathrm{c}}$] Peak-to-peak line profile asymmetry
\end{list}
\end{minipage}
\end{table}

\begin{table}
\begin{minipage}[t]{\columnwidth}
\caption{Flux ratios of models \#1 and \#2}
\label{table:flux-ratios}
\centering
\renewcommand{\footnoterule}{}
\begin{tabular}{c c c c c}
\hline\hline
Model & $i$ & $F_\mathrm{dc}/F_{*}^\mathrm{a}$ & $F_\mathrm{de}/F_{*}^\mathrm{b}$ & $F_\mathrm{dc+de}/F_{*}^\mathrm{c}$ \\
\hline
\#1  & $20^{\circ}$ & 5.13 & 2.78 & 7.92 \\
   & $60^{\circ}$ & 2.88 & 2.78 & 5.66 \\
\hline
\#2  & $20^{\circ}$ & 4.05 & 1.75 &  5.8 \\
   & $60^{\circ}$ & 2.45 & 1.75 &  4.2 \\
\hline
\end{tabular}
\begin{list}{}{}
\item[$^{\mathrm{a}}$] Disk interior to stellar flux ratio
\item[$^{\mathrm{b}}$] Disk edge to stellar flux ratio
\item[$^{\mathrm{c}}$] Disk interior + disk edge to stellar flux ratio
\end{list}
\end{minipage}
\end{table}

Figure \ref{fig:V1-0P10} shows the line profiles emerging from an unperturbed (dashed line) and an eccentric disk (solid line), assuming $20^{\circ}$ and $60^{\circ}$ disk inclination angles. We performed the synthetic spectral calculations in two disk models with different disk sizes, using the same hydrodynamical output of the 600th binary orbit. The computational domain of the synthetic spectral calculation was $0.4\,\mathrm{AU}\leq R \leq 40\,\mathrm{AU}$ and $0.2\,\mathrm{AU}\leq R \leq 20\,\mathrm{AU}$ for model\,\#1 and model\,\#2, where the binary separation is $40\,\mathrm{AU}$ and  $20\,\mathrm{AU}$, respectively (left and right panels of Fig.\,\ref{fig:V1-0P10}). In this way, the disk edge is at 0.4\,AU and 0.2\,AU for models \#1 and \#2, respectively. As the CO fundamental band is excited out to 2-3\,AU \citep{Najitaetal2007}, the disk material beyond 3\,AU does not contribute to the CO line emission in our model. However, material beyond this radii out to $\sim5\,\mathrm{AU}$ must be considered for the line-over-continuum calculation, since the dust in the superheated disk atmosphere contributes considerably to the continuum at $4.7\mathrm{\mu m}$. 

It can be clearly seen that the line profiles formed in an eccentric disk (solid lines) are asymmetric, assuming that the semimajor axis of the elliptic disk is perpendicular to or at least not aligned with the line-of-sight. The line-to-continuum is stronger in model \#2 than in model \#1, owing to the disk continuum (disk + disk edge) of model \#1 exceeding that of model \#2 (see details of the calculated fluxes in Table \ref{table:flux-ratios}) and the additional hot CO emission orbiting between 0.2\,AU and 0.4\,AU in model \#2. The depressed disk continuum for model \#2 may be due to: (1) the rim flux being smaller for a rim closer to the star, since the disk rim flux is $F_\mathrm{rim}\sim T_\mathrm{edge}^4R_\mathrm{edge}^2$, where the $T_\mathrm{edge}\sim R^{-2/5}$ (see, e.g., Appendix A of \citet{Regalyetal2010}), resulting in $F_\mathrm{rim}\sim R^{2/5}$; (2) the radial extension of the region close to the rim where the disk interior is irradiated directly by the rim (see, e.g. Appendix B of \citet{Regalyetal2010}) is larger for model \#1, resulting in a stronger continuum for model \#1 than model \#2.

The profile asymmetry does not differ significantly between the two models. Calculating the peak-to-peak line profile asymmetry as $A_\mathrm{pp}=|I_\mathrm{b}-I_\mathrm{r}|/(0.5[I_\mathrm{b}+I_\mathrm{r}]-1)$, where $I_\mathrm{r}$ and $I_\mathrm{b}$ are the continuum normalized line fluxes at the red and blue peaks, respectively, the asymmetry is $A_\mathrm{pp}\simeq20\%$ for all models (see details in Table \ref{table:lp-asymmetry}). After closer inspection of Fig.\,\ref{fig:V1-0P10}, the line center is clearly seen to shift toward the peak, which is in excess. The magnitude of line center shift also depends on the disk inclination angle and the disk extension. Obviously, the larger the inclination angle, the larger the line center shift. Moreover, the magnitude of the line center shift is larger in model \#2 than in model \#1 because of the additional emission of hot gas parcels on closer orbits in model \#2.

\subsection{Formation and variability of line profile asymmetry}

\begin{figure*}
	\centering
	\includegraphics[width=18cm]{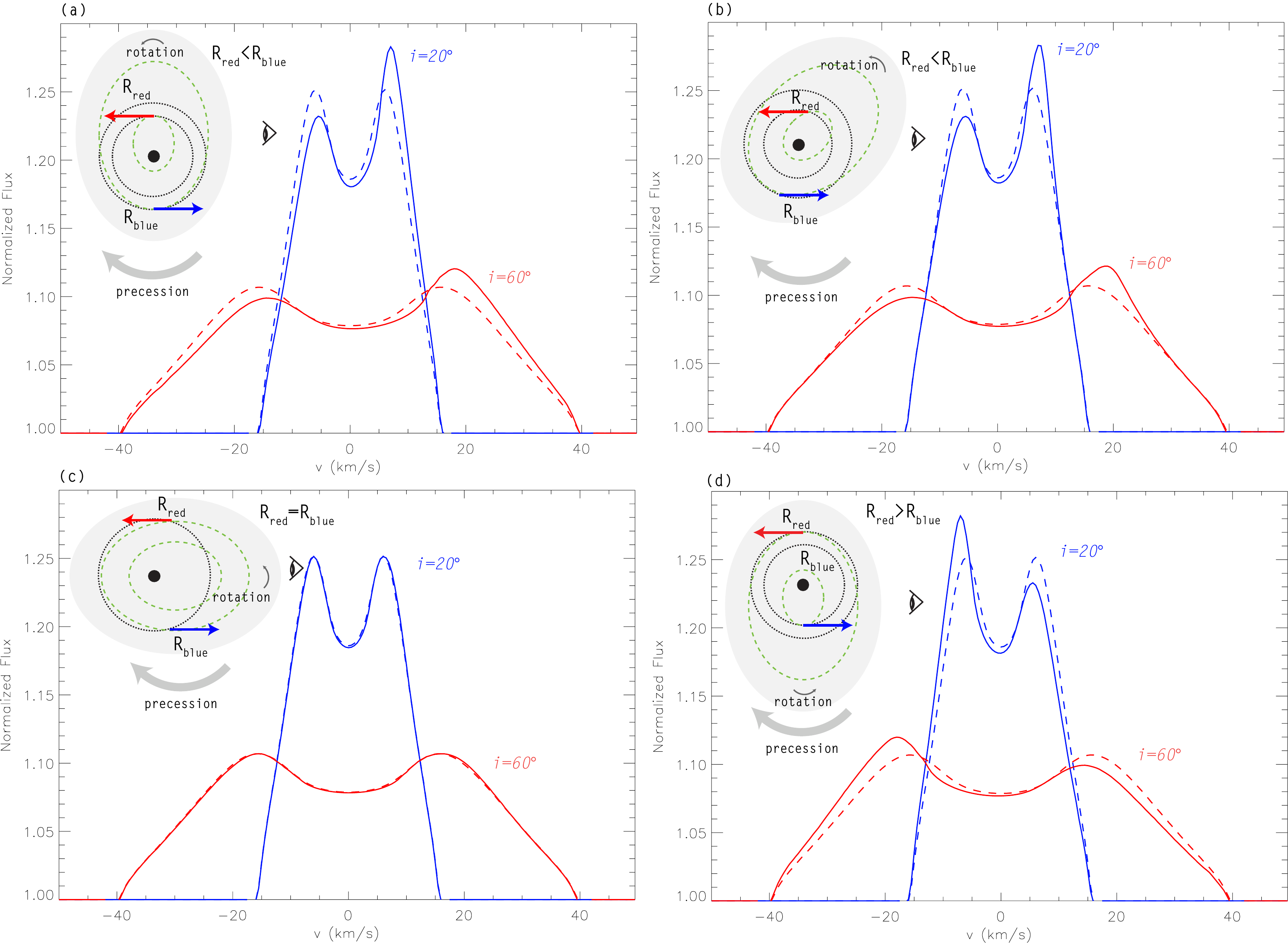}
	\caption{$V=1\rightarrow0$ P10 ro-vibrational CO line-profile variations during a full precession period of eccentric disk calculated in model \#1 after the 600th binary orbit. For comparison, the line profiles emerging in an unperturbed circularly Keplerian disk are shown with dashed lines. Small subfigures show the disk's apparent position with respect to the line of sight. The $i=20^{\circ}$ and $i=60^{\circ}$ inclination angles were assumed, shown in blue and red colors, respectively. The orbits of gas parcels are indicated by green dotted ellipses in the subfigures. \emph{Panels} {\bf a} and {\bf d}: show that the asymmetric line profiles emerging in the disk when seen normal to its semimajor axis. \emph{Panel} {\bf c}: shows that the symmetric line profiles emerging in the disk when seen parallel to its semimajor axis. \emph{Panel} {\bf b}: shows that the line profiles with a smaller amount of asymmetry emerging from a disk with intermediate position angle.}
	\label{fig:V1-0P10_var} 
\end{figure*}

The line profile distortion in eccentric disks occurs owing to the same phenomena as for the giant planet-bearing disks presented in \citet{Regalyetal2010}. But here the {\it whole disk is eccentric}, which allows us to give an even simpler explanation of the origin of asymmetry. This and the consequences of disk precession are explained in this section. 

To provide insight into the cause of the line profile variations, we calculated the CO line profiles during a full disk-precession period in model \#1. Part of the results are shown in Fig.\,\ref{fig:V1-0P10_var}. The line profile asymmetry is most prominent when the disk semimajor axis is seen perpendicular to the line-of-sight (Fig.\,\ref{fig:V1-0P10_var}, panel a and d). In contrast, the line profile asymmetry completely vanishes when the disk semimajor axis is seen parallel to the line-of-sight (Fig.\,\ref{fig:V1-0P10_var}, panel c). 

The formation of the asymmetry can be explained by the temperature difference of the gas parcels with the same absolute values of receding and approaching velocities. In an elliptic orbit, the Y velocity component at cylindrical coordinates $(R,\phi)$ can be given by
\begin{equation}
	V_\mathrm{Y}(R,\phi)=V_\mathrm{K}(R)\frac{\cos(\phi)+e}{\sqrt{1+e\cos(\phi)}},
	\label{eq:vel_Y_ecc}
\end{equation}
where $e$ is the orbital eccentricity of the gas parcels. For simplicity, we assume that the disk eccentricity is constant with radius. Gas parcels at the opposite side of the disk (one at apastron $R_\mathrm{ap}$ and the other at periastron $R_\mathrm{per}$), have the same absolute value of the velocity along the line-of-sight ($V_\mathrm{Y}(R_\mathrm{ap},\phi)=-V_\mathrm{Y}(R_\mathrm{per},\phi)$) if the equality
\begin{equation}
	R_\mathrm{ap}=\frac{1-e}{1+e}R_\mathrm{per}
	\label{eq:Rap-Rper}
\end{equation}
holds, using Eq. (\ref{eq:vel_Y_ecc}). We note that gas parcels at $R_\mathrm{ap}$ and $R_\mathrm{per}$ are not on the same orbit. Here we align the cylindrical coordinate system such that at periastron $\cos(\phi)=1$, and at apastron $\cos(\phi)=-1$. According to Eq. (\ref{eq:Rap-Rper}), $R_\mathrm{ap}<R_\mathrm{per}$ holds because the eccentricity is given by $0<e<1$ for elliptic orbits. As the gas parcels at apastron and periastron are approaching and receding, they contribute to the blue and red peaks, respectively. The disk is seen by the observer as shown in the subfigure of panel a of Fig.\,\ref{fig:V1-0P10_var}. Setting the radii of the gas parcels that contribute to the red and blue peaks to be $R_\mathrm{red}\equiv R_\mathrm{ap}$ and $R_\mathrm{blue}\equiv R_\mathrm{per}$, respectively, we find that $R_\mathrm{red}<R_\mathrm{blue}$. The temperature of the gas parcels at apastron exceeds that at periastron owing to the $T_\mathrm{atm}(R)\sim R^{-2/5}$ dependence of the atmospheric temperature on distance in the double-layer disk model. Thus, the gas parcels receding with speed $V$ make a larger contribution to the red peak than the gas parcels approaching with speed $-V$ to the blue peak (Fig.\,\ref{fig:V1-0P10_var}, panel a). 

Having precessed the disk by $45^{\circ}$, the red peak remains in excess, since although the expression Eq. (\ref{eq:Rap-Rper}) does not hold, $R_\mathrm{red}<R_\mathrm{blue}$ (Fig.\,\ref{fig:V1-0P10_var}, panel b). Although the magnitude of peak asymmetry does not change significantly, the departure of the line profile shape from the Keplerian one is smaller in the line wings.

By the time the disk has precessed by $90^{\circ}$, the distances of the gas parcels receding and approaching with the same speed are equal, thus $R_\mathrm{red}=R_\mathrm{blue}$. Obviously, their contributions to the red and blue peaks are just the same, resulting in no asymmetry in the line profile (Fig.\,\ref{fig:V1-0P10_var}, panel c). 

The disk is seen again perpendicular to its semimajor axis in a subsequent $90^{\circ}$ of disk precession (Fig.\,\ref{fig:V1-0P10_var} panel d). In this phase, the gas parcels at periastron and apastron are now  receding and approaching, respectively. In this case, $R_\mathrm{red}\equiv R_\mathrm{per}$ and $R_\mathrm{blue}\equiv R_\mathrm{ap}$, thus the distance relation of the receding and approaching gas parcels is found to be $R_\mathrm{blue}<R_\mathrm{red}$ using Eq. (\ref{eq:Rap-Rper}). The level of red-blue peak asymmetry is nearly the same for disks seen at antiparallel position angles perpendicular to the semimajor axis (Fig.\,\ref{fig:V1-0P10_var}, panel a and d). Although the disk eccentricity varies during the disk precession (Fig.\,\ref{fig:profile_evol}, panel d), this can be explained by its amplitude being small in the regions $R<2-3\,\mathrm{AU}$ where the CO is excited.

\begin{figure}
	\centering
	\includegraphics[width=\columnwidth]{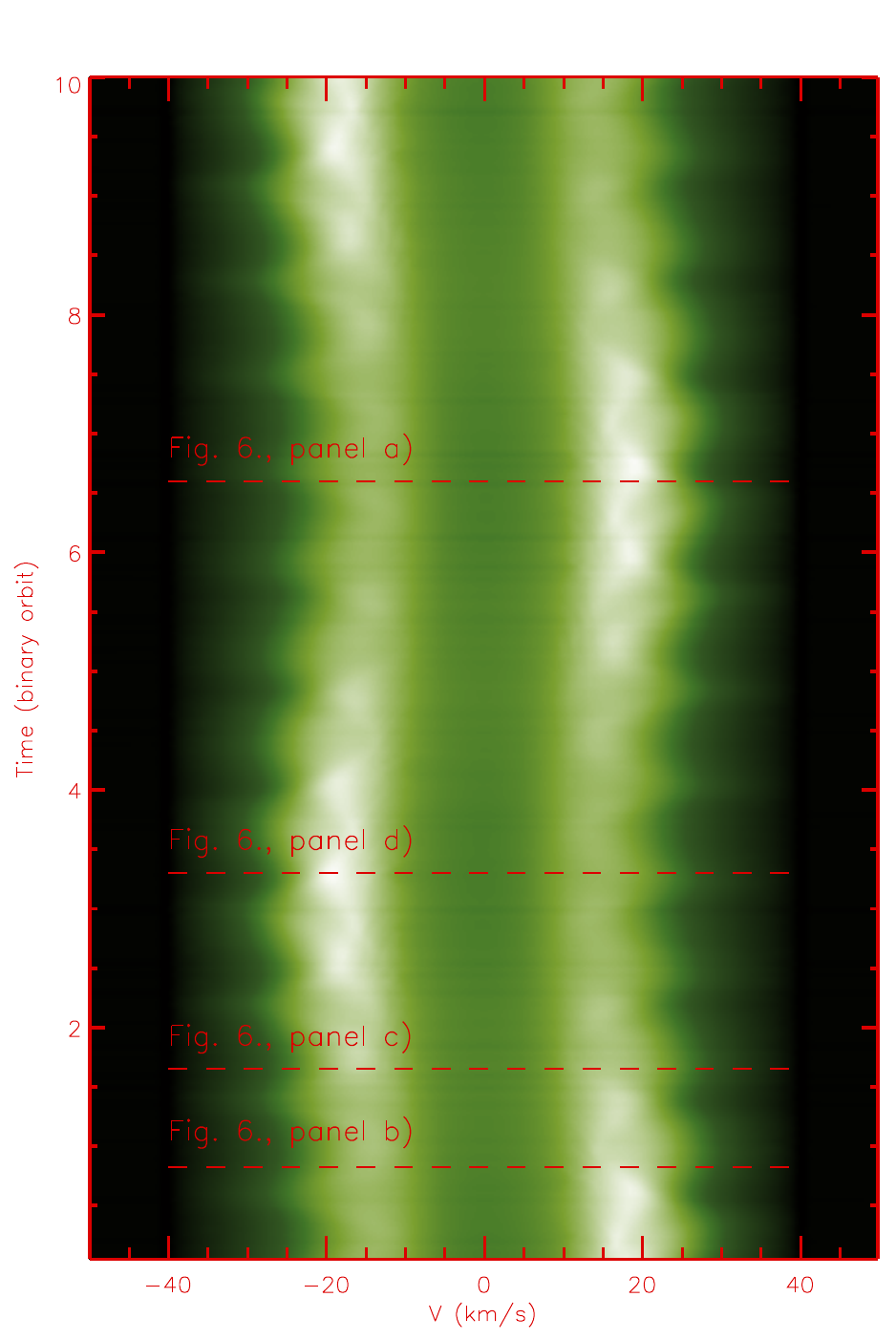}
	\caption{Trailed spectra of $V=1\rightarrow0$ P10 ro-vibrational CO line profile at $4.7\mu$m emerging from the quasi-steady eccentric disk state in model \#1. The line profiles were calculated with cadence of 1/20 binary orbit, and the trailed spectra covers 10 binary orbits. Epochs of the line profiles shown in Fig.\,\ref{fig:V1-0P10_var} are marked with dashed lines.}
	\label{fig:ts} 
\end{figure}

To illustrate the line profile variations during 200 binary orbits, we present trailed spectra of CO in Fig.\,\ref{fig:ts}. The periodic change in the red-blue peak asymmetry caused by the disk precession is clearly visible. In addition, smaller variations in the line wings with periods $\sim0.7$ times the binary orbital period are also visible. This can be explained by small-magnitude disk eccentricity variations owing to its dependence on the apparent position angle of the secondary with respect to the disk semimajor axis. These variations are similar to those that occur in superhump binaries, where an additional period is present as a beat period between the precessing disk and the binary orbit. However, in the superhump case, the period is slightly longer than the binary period because of the prograde disk precession \citep{GoodchildOgilvie2006}. In our case, the period is shorter than the binary orbital period because the disk precession is retrograde.

To summarize, the line profile becomes asymmetric in the quasi-steady eccentric disk state. The period of line profile asymmetry is equal to that of the disk precession. Two asymmetric phases alternate with each other, and there are two symmetric phases amongst them. However, the variation in the red-blue peak asymmetry is unlikely to be observed within a decade because of the long precession period, which is $\sim250\,\mathrm{yr}$ for a 40\,AU separation binary. Nevertheless, as the line wings are also subject to variations of period $\sim0.7$ times that of the binary period, it might be detectable for $\sim10\,\mathrm{AU}$ separation binaries with $\sim2\,M_\odot$ primaries.

\section{Parameter study}

To investigate under which conditions the disk will develop eccentricity, we performed an extensive hydrodynamical study. The following parameters were varied within physically reasonable intervals: the binary mass ratio $q$, the viscosity parameter $\alpha$, the binary eccentricity $e_\mathrm{bin}$, the disk aspect ratio $h$, the flaring index $\gamma$, and the disk-to-secondary mass ratio $q_\mathrm{disk/sec}$. In addition, we investigated the effects of the open outflow and the rigid boundary conditions at the inner boundary of the disk.

To speed up our calculations, we assumed a larger disk inner radius ($R_\mathrm{in} = 0.05 a_\mathrm{bin}$) in the parameter study than used previously ($R_\mathrm{in} = 0.01 a_\mathrm{bin}$) in Sect. 2. We ran several models with $0.01 a_\mathrm{bin}\leq R_\mathrm{in}\leq0.05 a_\mathrm{bin}$. The results demonstrated that neither the disk eccentricity profile nor the evolution of the average disk eccentricity depends significantly on the choice of the disk inner radius. The hydrodynamical simulations performed in this parameter study cannot be used to calculate the strength of the CO emission lines, because when assuming that the distance unit is either 40\,AU or 20\,AU, the hot emitting inner part of the disk is not involved in the calculation. This does not prevent the evolution of the eccentric quasi-steady state of the disk being able to be studied and described in detail.

\subsection{Results for the disk eccentricity and radius}

\begin{figure*}
	\centering
	\includegraphics[width=\textwidth]{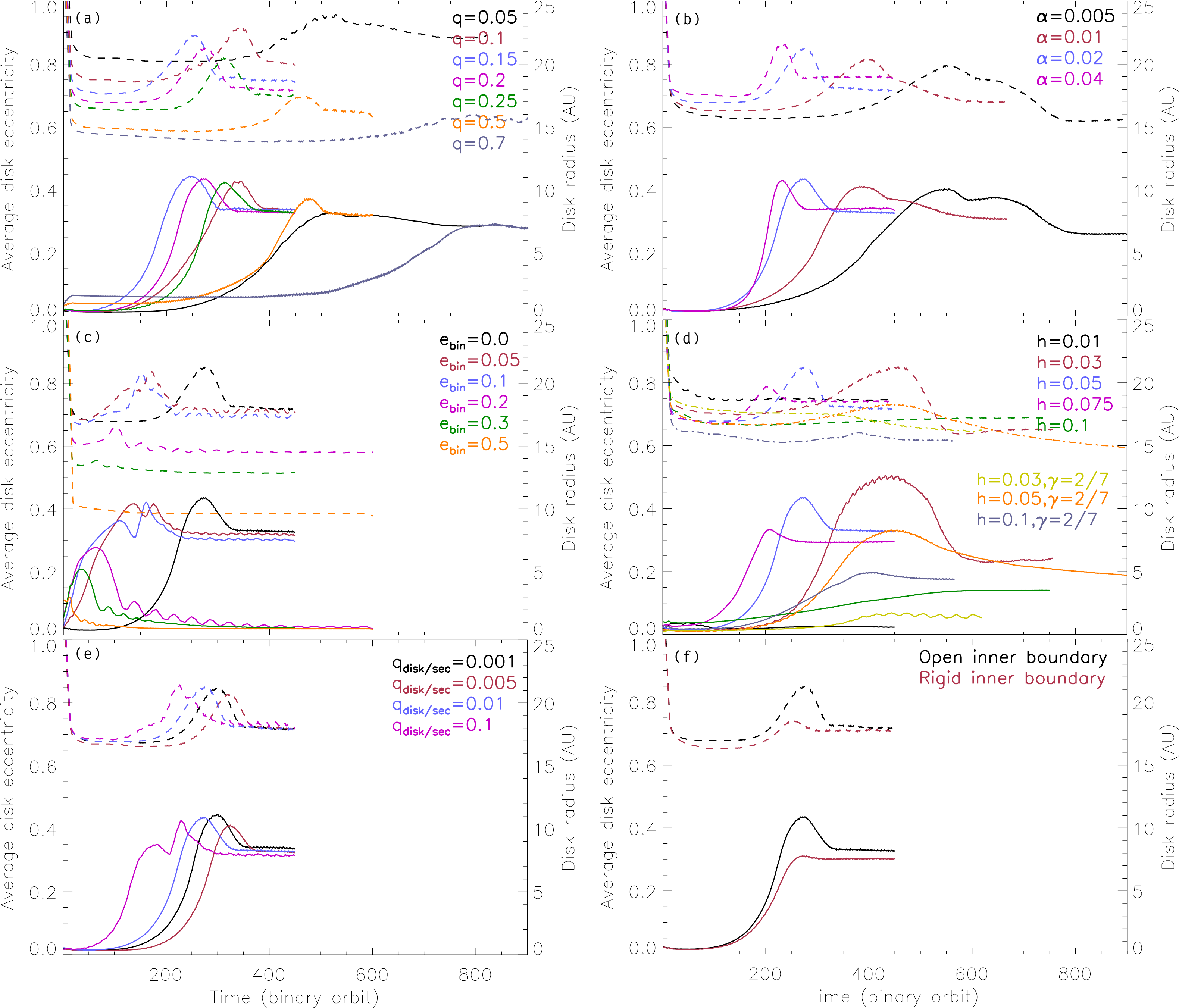}
	\caption{Evolution of the disk radii (dashed curves) and the average disk eccentricities (solid curves) for different models in our parameter study: \emph{Panel {\bf a}}) models with binary mass ratios  $0.05\leq q\leq0.7$, viscosity parameter $\alpha=0.02$, and aspect ratio $h=0.05$; \emph{Panel {\bf b}}) models with $0.005\leq\alpha\leq0.4$, $q=0.2$, and $h=0.05$; \emph{Panel {\bf c}}) models with binary orbital eccentricity $0.0\leq e_\mathrm{bin}\leq0.5$, $q=0.2$, $\alpha=0.02$, and $h=0.05$; \emph{Panel {\bf d}}) models with $0.01\leq h\leq0.1$, $q=0.2$, and $\alpha=0.02$; flaring disk models with $\gamma=2/7$ are shown with dot-dashed curves; \emph{Panel {\bf e}}) models with disk mass $0.001\leq q_\mathrm{disk/sec}\leq0.1$, $q=0.2$, $\alpha=0.02$, and $h=0.05$; \emph{Panel {\bf f}}) models with open outflow and rigid boundary conditions assuming $q=0.2$, $\alpha=0.02$, and $h=0.05$.}
	\label{fig:ecc_r_multi} 
\end{figure*}

The radial extension of a circumprimary disk is determined by the tidal truncation, whose efficiency depends on the binary mass ratio and possibly other physical parameters of the disk itself such as the magnitude of the disk viscosity. To characterize the time evolution of the disk size, we defined the disk radius as the distance measured from the primary star containing 90\% of the disk material. The disk eccentricity is calculated at each individual grid cells as mentioned in Sect. 2.1. To characterize the disk eccentric state, the eccentricity distribution was azimuthally and  within the disk radius radially averaged. Since the 3:1 Lindblad resonance point -- which is responsible for the eccentricity excitation \citep{Lubow1991} -- is close to the outer disk edge, the disk eccentricity is strongly connected to the disk radius evolution. Therefore, it is crucial to describe the development of the disk radius and the average disk eccentricity in parallel.

The evolution of the average disk eccentricity tends to follow the evolution of the disk radius. The disk is tidally truncated after a few tens of binary orbits at a radius $R\simeq 0.35 - 0.54 a_\mathrm{bin}$ for all disk models. Later on, the disk increases in size, then shrinks, and finally approaches a size of $R\simeq0.35-0.4a_\mathrm{bin}$, when the quasi-steady eccentric state has been reached. The disk remains circular for a couple of hundred orbits, until at some point the average disk eccentricity increases abruptly reaching a temporary maximum, which is followed by a decrease. After this jump, the eccentricity curves retain their constant values, no further jumps or sudden changes occurring within the simulation time.

The evolution of the disk radius and average disk eccentricity differs fundamentally from the above-described general cases for thin ($h = 0.01$) and thick ($h = 0.1$) models, respectively. While in a thick model, the disk does not expand temporarily and the average disk eccentricity remains low, for thin models the disk eccentricity does not develop at all. Moreover, for significant binary eccentricity ($e_\mathrm{bin}\geq0.2$), although the disk expands and becomes eccentric temporarily, the average disk eccentricity decays within $\sim100$ subsequent binary orbits. 

The final disk radius is significantly affected by the mass ratio $q$ of the binary system in models with $0.05\leq q \leq 0.7$ and $\alpha=0.02$. We found that the larger the mass ratio, the smaller the final disk radius (Fig.\,\ref{fig:ecc_r_multi}, panel a, dashed curves). The final disk radius is in the interval $0.35 - 0.45 a_\mathrm{bin}$, which corresponds to 14-18\,AU assuming $a_\mathrm{bin}=40\,\mathrm{AU}$ for binary separation. On the other hand, the magnitude of the final average disk eccentricity does not depend on $q$ (Fig.\,\ref{fig:ecc_r_multi}, panel a, solid curves). We also found that the azimuthally averaged eccentricity profile are practically the same in the quasi-steady disk state for all models. The timescale of the disk eccentricity evolution shows no clear dependence on $q$. The eccentricity growth rate  increases monotonically with increasing binary mass ratio in the range of $0.05\leq q\leq 0.15$, while for $q\geq 0.2$ the growth rate decreases with increasing $q$.

To investigate the effect of the magnitude of the viscosity on the disk eccentricity evolution, we performed simulations with $0.005 \leq \alpha \leq 0.04$ and $q=0.2$. We found that with increasing viscosity, the disk reaches a quasi-steady eccentric state at larger disk radius. The final disk radii are in the range of $0.33a_\mathrm{bin}-0.42a_\mathrm{bin}$, corresponding to 13.5\,AU-17\,AU (Fig.\,\ref{fig:ecc_r_multi}, panel b, dashed curves). The magnitude of the viscosity, however, has no strong effect on the final value of the average disk eccentricity (Fig.\,\ref{fig:ecc_r_multi}, panel b, solid curves). Although the final azimuthally averaged disk eccentricity profiles are also very similar to each other, a slightly lower value of the final average disk eccentricity was found for the $\alpha=0.005$ low viscosity model. On the other hand, the time required to reach the quasi-steady eccentric disk state decreases with increasing values of $\alpha$. 

To investigate the effect of the initial orbital eccentricity of the binary on the disk eccentricity evolution, we performed simulations in which the secondary was placed on an initially eccentric orbit with the same apastron distance as used for circular models assuming $q=0.2,\alpha=0.02$. The initial eccentricity $e_\mathrm{bin}$ was in the range of 0.05-0.5. We found that the final disk radius is also sensitive to the binary eccentricity. As a general tendency, the final disk radius for an eccentric binary falls short of that of the circular binary case (Fig.\,\ref{fig:ecc_r_multi}, panel c, dashed curves). The final disk radius decreases with increasing binary eccentricity above $e_\mathrm{bin}=0.2$. The disk becomes permanently eccentric only for $e_\mathrm{bin}\leq0.2$ models, in which the final value of the average disk eccentricity is nearly the same (Fig.\,\ref{fig:ecc_r_multi}, panel c, solid curves). For higher binary eccentricity, the disk becomes only temporarily eccentric. There is no permanent eccentric disk state at all in $e_\mathrm{bin}\geq0.2$ models. The lifetime of the temporary eccentric disk state decreases significantly with increasing binary eccentricity above $e_\mathrm{bin}=0.2$. 

We modeled circumstellar disks with various aspect ratios in the interval $0.01 < h < 0.1$ with $q=0.2$ and $\alpha=0.02$. We found that the choice of the disk aspect ratio substantially affects the final disk radius (Fig.\,\ref{fig:ecc_r_multi}, panel d, dashed curves), which increases with increasing $h$ for models $0.03\leq h\leq0.05$, and decreases for thicker disks ($0.075\leq h\leq0.1$). For thick ($h=0.1$) and thin disks ($h=0.01$), the disk radii are stabilized early (no temporary disk puff-up being present) at $\sim$17\,AU and $\sim18$\,AU, respectively. The final average disk eccentricity displays a similar dependence on the aspect ratio as the disk radius (Fig.\,\ref{fig:ecc_r_multi}, panel d, solid curves). The final value of average disk eccentricity is at its largest for the $h=0.05$ model, but for either larger or smaller aspect ratios the disk eccentricity profile is somewhat flatter. We note, however, that although the average disk eccentricity is lower for $h\geq0.075$ than for $h=0.05$, the azimuthally averaged eccentricity profile of the inner disk ($R<4\,\mathrm{AU}$) is similar. For a very thick disk ($h=0.1$), we found that there is no temporary high eccentric state, and the disk eccentricity is significantly lower than in the thinner models. We emphasize that disk eccentricity does not develop at all in the very thin disks ($h=0.01$) case. 

We also studied the eccentricity evolution for the flaring disk approximation, assuming a $\gamma = 2/7$ flaring index \citep{ChiangGoldreich1997}. The disk aspect ratio ($h(R)=h(a_\mathrm{bin})R^\gamma$) was $h(a_\mathrm{bin})=0.03-0.1$ at the distance of the secondary. In these models, the disk radius is stabilized at lower values than in the corresponding non-flaring models (Fig.\,\ref{fig:ecc_r_multi}, panel d, dot-dashed curves). The average disk eccentricity remains below that of the appropriate non-flaring ones, except for the thick $h=0.1$ disk case, where the disk eccentricity of the flaring model exceeds the appropriate non-flaring one (Fig.\,\ref{fig:ecc_r_multi}, panel d, dot-dot-dashed curves). Noteworthy is the significantly slower eccentricity growth rate for flaring models. In general, the azimuthally averaged disk eccentricity profiles of flared models stay below the corresponding non-flaring models, except for the thick ($h=0.1$) model, where the flared eccentricity profile is above the corresponding non-flaring one.

The hydrodynamical equations do not depend on the disk mass since $\Sigma_0$ drops out of the hydrodynamical equations \citep{Kley1999}. This is true as long as we use the isotherm equation of state ($p=\Sigma c_\mathrm{s}^2$) and neglect the change in the binary orbital parameters. The gravitational potential of the secondary that perturbs the disk, however, is affected by the orbital parameters of the secondary, which might be subject to change. The magnitude of the torque exerted by the disk on the secondary indeed depends on the disk mass. Thus, the final dynamical state of the disk might be influenced by the disk-to-secondary mass ratio $q_\mathrm{disk/sec}$. To investigate whether the disk mass influences the disk eccentricity evolution, we performed several simulations assuming a $0.0002 - 0.02\,M_{\odot}$ disk mass with $q=0.2$ and $\alpha = 0.02$, corresponding to $0.001\leq q_\mathrm{disk/sec}\leq0.1$. We found that the final disk radius does not depend on the initial disk mass (Fig.\,\ref{fig:ecc_r_multi}, panel e, dashed curves). Similarly, the final value of the average disk eccentricity is also insensitive to the disk mass (Fig.\,\ref{fig:ecc_r_multi}, panel e, solid curves). However, we concluded that the disk eccentricity evolution tends to begin at earlier epochs for the more massive disks. We note that the azimuthally averaged disk-eccentricity profile is practically independent of the disk mass being in the above-mentioned range. This can be explained by the changes in the orbital parameters of the secondary being negligible, i.e., $\Delta a/a_{in}<0.001$ and $\Delta e_\mathrm{bin}<0.001$.

Finally, we investigated whether the choice of the inner boundary influences the disk eccentricity evolution. This is an important question because a feasible inner boundary condition might differ when the disk inner rim is close to or far from the primary. There are two extreme boundary conditions regarding the outflow mass flux at the inner disk edge: maximum allowed and minimum (zero) mass flux occurs for open and rigid-wall boundary conditions, respectively. An open boundary condition, in which case the mass accretion rate at the inner edge is not influenced by the star, seems to be feasible for cases when the inner disk is farther away from the star, e.g. when the inner disk is cleared by photoevaporation, for example \citet{Shuetal1993} and \citet{Clarkeetal2001}. When the disk inner edge is close to the stellar surface (i.e.  at the magnetospheric radius) the radial velocity component of the migrating material practically vanishes owing to the equilibrium of the stellar magnetospheric pressure and the ram pressure of accreting material \citep{Koenigl1991}. In an extreme situation, the accretion ceases, thus a perfectly rigid wall assumption might be valid. Thus, we applied an open outflow (where both the azimuthal and the radial components of the velocity are the same in the ghost cells) and rigid inner boundary (at which the radial velocity component changes its sign without changing its magnitude in the ghost cells) conditions for model $\alpha=0.02,q=0.2$. Comparing the results of models with open and rigid inner boundary conditions, one can see that both the final disk radius and final averaged disk eccentricity profiles are independent of the choice of inner boundary condition (Fig.\,\ref{fig:ecc_r_multi}, panel f, dashed curves). The time needed to reach the quasi-steady eccentric disk state is also similar for both boundary conditions. Noteworthy is the difference in the magnitude of averaged disk eccentricity value during the temporarily high eccentricity states.

\section{Discussion}

In his pioneering work, \citet{Lubow1991,Lubow1991b} showed that tidally induced eccentricity develops in circumprimary disks of close-separation binaries because of the effect of the 3:1 eccentric inner Lindblad resonance. \citet{Kleyetal2008} numerically investigated the eccentricity development in circumprimary disks using spatially uniform kinematic viscosity. \citet{Lubow2010} then explored the eccentricity growth rate by linear eccentricity evolution formulated by \citet{GoodchildOgilvie2006} neglecting the effect of viscosity. In our approach, however, the viscosity of the disk material is provided by the turbulence within the disk \citep{ShakuraSunyaev1973}, where the kinematic viscosity increases with increasing stellar distance. In what follows, we compare our results to that of \citet{Kleyetal2008} and \citet{Lubow2010}.

\subsection{Comparison of simulations}

In line with these studies, our simulations demonstrate that the circumprimary disk becomes eccentric with final average eccentricity of $\bar{e}\simeq0.25$, assuming $\alpha$-type viscosity in the range $0.005\leq\alpha\leq0.04$. In good accordance with the results of \citet{Kleyetal2008}, we found that the disk's final average eccentricity does not depend on the binary mass ratio. In addition to the results of \citet{Kleyetal2008}, we confirmed these findings for mass ratios higher than 0.3. However, the time required to reach the quasi-steady eccentric disk state was found to decrease with increasing mass ratio only up to $q=0.15$. Above this mass ratio, the disk eccentricity evolution decelerated significantly, and ever more slowly with increasing $q$ in our simulations. \citet{Kleyetal2008} found that the increase in the eccentricity growth rate is continuous up to $q=0.3$. This behavior of eccentricity growth rate can be explained if tidal forces lower the disk density near the resonance, where the eccentricity is excited, and the eccentricity growth rate then is reduced \citep{Lubow2010}.

\citet{Lubow2010} qualitatively expected that the eccentricity growth would accelerate as the viscosity increased because the disk is then more likely to spread outwards. Along with the results of \citet{Kleyetal2008}, we confirmed this prediction as the disk eccentricity growth rate is found to monotonically increase with $\alpha$.

\citet{Kleyetal2008} found that the disk final eccentricity decreases with decreasing values of the dimensionless kinematic viscosity up to $\nu_\mathrm{k}\leq 10^{-5}$, while below this value the disk eccentricity is found to be saturated (Fig.\,8. of \citet{Kleyetal2008}). In the $\alpha$-type viscosity approach, the kinematic viscosity is $\nu_{k}=\alpha h^2R^{1/2}$. Hence, in our models the dimensionless kinematic viscosity is $2.5\times10^{-6}<\nu_\mathrm{k}<6.5\times10^{-5}$ within the disk radius assuming  that $0.005\leq\alpha\leq0.04$, and a $h=0.05$ non-flaring disk. Thus, the values of the kinematic viscosity modeled in our simulations are close to the range where the disk eccentricity is found to be independent of viscosity by \citet{Kleyetal2008}. 

The evolution of the disk eccentricity departs significantly from the general trends for eccentric binaries with $e_\mathrm{bin}\geq0.2$. The 3:1 Lindblad resonance responsible for the disk eccentricity \citep{Lubow1991} lies farther toward the primary, and the disk is truncated at smaller radius in an eccentric binary than in circular ones. As a consequence, the excitation of 3:1 Lindblad resonance might be inefficient, resulting in only a temporary eccentric state with maximum value of $\bar{e}_\mathrm{disk}<0.3$ that lasts for only a hundred binary orbits.

We have explored a slightly wider aspect-ratio range than \citet{Kleyetal2008}. In their investigation, the disk eccentricity displays a weaker linear increase, the aspect ratio being in the range of 0.02-0.06. \citet{Lubow2010} concluded that the eccentricity growth rate increases with increasing sound speed, being equivalent to the disk geometrical thickness. We have confirmed this result. However, for very thick disks ($h\geq0.75$) both the final disk eccentricity and the growth rate were found to be smaller than for the $h=0.05$ models, meaning that the aspect ratio-eccentricity relation turns above $h=0.75$. In addition, the disk eccentricity evolution is found to be inhibited for very thin disks ($h=0.01$). To explain this odd relation between the disk geometrical thickness and eccentricity development, further theoretical investigation will be required in which the effect of the disk viscosity can also taken into account.

Noteworthy is the disk eccentricity evolution in flared disk models, namely, we have found that a significantly lower average disk eccentricity develops in the case of a flaring disk geometry than in a flat one. However, there is an exception, i.e., the $h(a_\mathrm{bin)}=0.1$ thick flared model, where the resulting disk eccentricity is slightly above the corresponding non-flaring one. This may occur because a flaring disk is always thinner on average than the corresponding non-flaring one in our models. Thus, flaring disks are thinner at small $R$, where they are hence similar to the non-flaring models with small $h$, which also do not have high eccentricity.

\subsection{Observability of the line profile asymmetry}

We have shown that the eccentric circumprimary disk of a binary, containing considerable amount of gas, produces asymmetric CO ro-vibrational line profiles because of the eccentric orbits ($\bar{e}_{disk}\simeq0.2$) of the CO emitting gas parcels. The largest line-profile asymmetry measured between the red and blue peaks of the $V=1\rightarrow0$\,P(10) transition is $\sim20-25\%$, depending on the disk inclination angle. However, we keep in mind that the magnitude of the line profile asymmetry depends on the position angle of the eccentric disk on the sky with respect to the line-of-sight. To observe the largest line profile asymmetry, the disk must be aligned perpendicular with its semimajor axis to the line-of-sight (Fig.\,\ref{fig:V1-0P10_var}, panel a or d). Taking this consideration into account, we now present a method to determine the disk eccentricity profile by fitting the CO line profile asymmetry.

\begin{figure}
	\centering
	\includegraphics[width=\columnwidth]{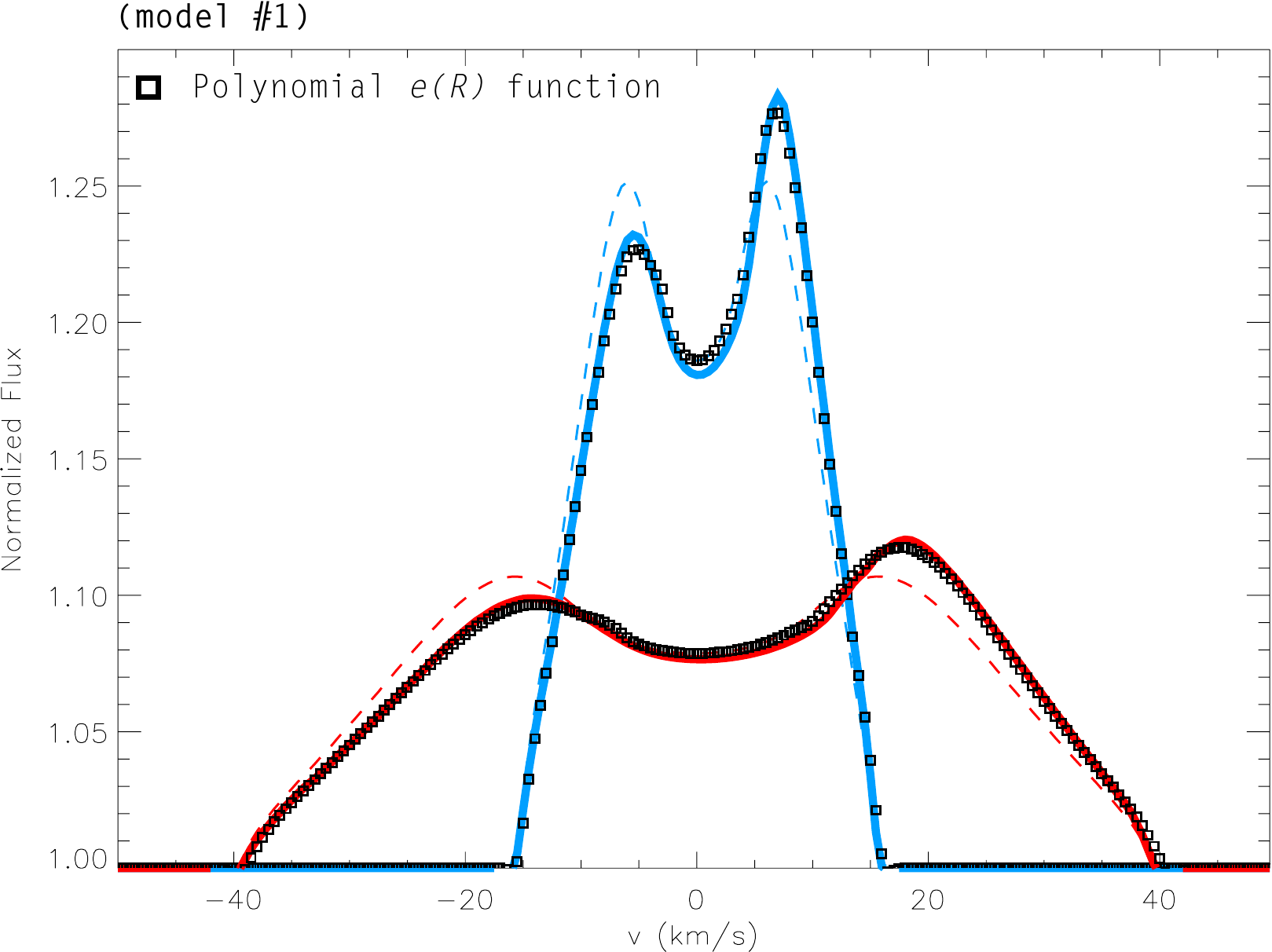}
	\caption{Line profiles formed in circular Keplerian disks (dashed curves), eccentric disks using the velocity distribution given by the hydrodynamical simulations (solid curves), and eccentric disk using the best-fit eccentricity profile (squares). The line profiles are calculated for disk inclinations $i=20^\circ$ (blue) and $i=60^\circ$ (red).}
	\label{fig:plynomial_ecc}
\end{figure}

Assuming that the disk plane is aligned with the binary orbital plane \citep{Moninetal2006}, which is measurable by the well-known astrometric technique \citep{Atkinson1966}, the disk inclination angle can be determined. The disk position angle on the sky can also be determined using the spectro-astrometric technique \citep{Porter2005}, which was applied successfully by \citet{Pontoppidanetal2008}. Knowing the disk inclination and the position angle of the eccentric disk on the sky, the model degeneracy can be resolved. Taking these considerations together, one can find the best-fit disk model in terms of the disk eccentricity, using our semi-analytical spectral model presented in \citet{Regalyetal2010}. The Doppler shift of the gas parcels at a given $R,\phi$ point in the eccentric disk seen perpendicular to its semimajor axis and inclined by $i$ can then be given by
\begin{eqnarray}
	\Delta\nu(R,\phi,i)&=&\frac{\nu_0}{c}\frac{V_\mathrm{K}(R)}{\sqrt{1+e(R)\cos(\phi)}}\times \nonumber\\
	&&\left\{\sin(\phi)+(\cos(\phi)+e(R))\right\}\sin(i),
	\label{eq:Doppler-shift-ecc}
\end{eqnarray}
where the disk eccentricity profile, $e(R)$, can be approximated by a quadratic polynomial function
\begin{equation}
	e_\mathrm{disk}(R)\simeq a_0+a_1 R+a_2R^2.
	\label{eq:aprox-ecc}
\end{equation}
Searching for a best-fit eccentricity profile for model \#1, we found that $a_0=-0.0385301$, $a_1=0.136924$, and $a_2=-0.0092516$. Figure \ref{fig:plynomial_ecc} shows the line profiles calculated in model \#1 using a circular Keplerian Doppler shift (solid curves) approximating the disk eccentricity using Eq. (\ref{eq:Doppler-shift-ecc}) (square symbols) befor calculating the Doppler shift. The line profiles emerging from the circular Keplerian disk are also shown (dashed curves). The line profiles are also calculated assuming $i=20^\circ$ (blue color) and $i=60^\circ$ (red color). One can see that the models using an approximated disk eccentricity (Eq. \ref{eq:aprox-ecc}) perfectly fit the line profiles calculated by the models using the hydrodynamical velocity distributions. Consequently, the disk eccentricity profile can be estimated by searching for the best-fit model of an asymmetric CO ro-vibrational line profile, with knowledge of both the disk position angle and the disk inclination angle. 

A fundamental uncertainty about the observability of eccentric circumprimary disks whether is there enough time to develop the disk eccentricity at all, before the disk is depleted in the visous timescale? Since the disk eccentricity evolution ceases within $\sim0.2\,\mathrm{Myr}$ in all of our models (Fig.\,\ref{fig:ecc_r_multi}), we expect that the eccentric state is occurred well within the disk lifetime. We search for the best-fit function of the evolution time required to reach the maximal eccentric state as a function of the viscosity in the form of $t_\mathrm{max}=a\alpha^b$, finding that $a=0.011$ and $b=-0.47$. Thus, within limitations of our modeling, the minimum viscosity required to develop an eccentric disk in a 40\,AU separation binary is $\alpha>6.8\times10^{-5},2.2\times10^{-6}$, and $5\times10^{-7}$ assuming 1\,Myr, 5\,Myr, and 10\,Myr gas depletion timescale, respectively.

\subsection{Disk thickness for T\,Tauris}

As we have shown, the circumprimary disk in a binary system with $e_\mathrm{bin}\leq0.2$ undergoes a significant increase in eccentricity for an aspect ratio in the range of $0.03\leq h \leq 0.075$, although there might be exceptional cases. The final average disk eccentricity is found to be significantly lower for thick circumprimary disks with aspect ratios of $h\geq0.1$ (Fig.\,\ref{fig:ecc_r_multi}, panel d). Moreover, for thin disks ($h\leq0.01$) the disk does not become eccentric at all. In what follows, we predict the thickness of the protoplanetary disk formed surrounding a T\,Tauri type star.

\begin{figure}
	\centering
	\includegraphics[width=\columnwidth]{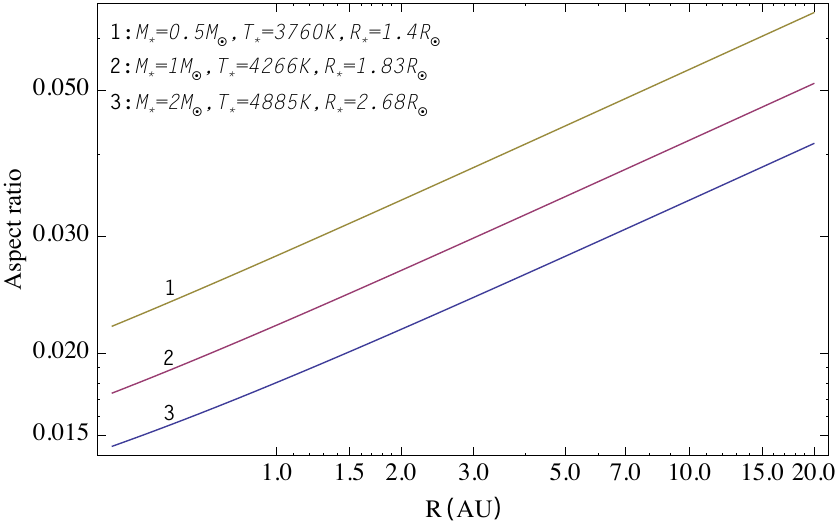}
	\caption{Disk aspect-ratio profiles calculated numerically using Eqs. (\ref{eq:aspect_ratio})-(\ref{eq:grazing-angle}) for three different mass T\,Tauri type stars. The stellar parameters assumed in the calculations are indicated.}
	\label{fig:aspect}
\end{figure}

We assume that the gas in the circumprimary disk is in vertical hydrodynamical equilibrium. Assuming that the disk is geometrically thin (i.e., $h(R)=H/R\gg1, R\gg R_*$), the disk aspect ratio can be given by
\begin{equation}
	h(R)=\sqrt{\frac{kT(R)R}{\mu_\mathrm{g}m_\mathrm{p}GM_*}},
	\label{eq:aspect_ratio}
\end{equation}
where $M_*$ is the stellar mass, $T(R)$ is the temperature profile of the disk, $k$ is the Boltzmann constant, $\mu_\mathrm{g}\simeq2.3$ is the mean molecular weight of the disk gas, and $m_\mathrm{p}$ is the proton mass (see e.g. \citet{DullemondDominik2004b}). In the flaring disk model of \citet{ChiangGoldreich1997}, the disk interior temperature is
\begin{equation}
	T(R)\simeq\left(\frac{\delta(R)}{4}\right)^{1/4}\left(\frac{R_*}{R}\right)^{1/2}T_*,
	\label{eq:Teff}
\end{equation}
assuming hydrostatic and radiative equilibrium. Here the accretion heating is neglected, i.e., only the stellar irradiation is taken into account. In Eq. (\ref{eq:Teff}), $\delta(R)$ is the grazing angle of stellar irradiation entering the disk atmosphere, which can be approximated as 
\begin{equation}
	\delta(R)\simeq\frac{2}{5}\frac{R_*}{R}+\frac{8}{7}\left(\frac{T_*}{T_c}\right)^{4/7}\left(\frac{R_*}{R}\right)^{-2/7},
	\label{eq:grazing-angle}
\end{equation}
where $T_\mathrm{g}=GM_*\mu_\mathrm{g}/kR_*$. For T\,Tauri stars, one can write $(T_*/T_\mathrm{g})^{4/7}\simeq0.007$. For an average T\,Tauri star, if $R\gtrsim0.1\,\mathrm{AU}$, $\delta(R)$ can be approximated by the second term of the right-hand side of Eq.  (\ref{eq:grazing-angle}), because this term dominates the equation. After some algebra, we get
\begin{equation}
	h(R)\simeq0.02 \left(\frac{R_*}{R_\odot}\right)^{3/14}\left(\frac{T_*}{T_\odot}\right)^{1/2}\left(\frac{M_*}{M_\odot}\right)^{-1/2}\left(\frac{R}{\mathrm{AU}}\right)^{4/7}.
	\label{eq:aspect_ratio_expl}
\end{equation}
Figure \ref{fig:aspect} shows the disk aspect ratio using Eq. (\ref{eq:aspect_ratio})-(\ref{eq:grazing-angle}) for three different masses of T\,Tauri type stars. (The stellar parameters were taken from \citet{Siessetal2000} assuming $2.5\,\mathrm{Myr}$ for the stellar age.) As one can see, the disk aspect ratio  increases with decreasing stellar mass, and is always above 0.01. Although the disk temperature is lower for lower mass stars, the squeezing effect of the stellar gravitational force is also smaller, resulting in  a higher disk aspect ratio. Since the disk aspect ratio is $\sim0.02-0.05$ on average, the disks around T\,Tauri type stars are thick enough to be eccentric. We note, however, that the SEDs of disks, mapping the dust rather than the gas disk around very low mass stars, can be fitted assuming significantly lower aspect ratios \citep{Szucsetal2010}. Accordingly, we expect the circumprimary disks encircling a T\,Tauri type young primary star in a binary system with $e_\mathrm{bin}\leq0.1$ to produce asymmetric line profiles owing to the development of disk eccentricity.

\subsection{Limits of our model}

In this study we have used a simple 2D disk model, although our assumptions are reasonable to investigate the effect of dynamically perturbed circumprimary disk on CO ro-vibrational line profiles. In the hydrodynamical calculations, we have neglected the disk self-gravity. Although the Toomre parameter \citep{Toomre1964} is $Q\gg 1$ in our models, the disk self-gravity might modify the picture. For global disturbances, such as significant disk ellipticity, even though the discs are gravitationally stable, pressure and self-gravity can be equally important \citep{Papaloizou2002}. In this case, the parameter measuring the importance of self-gravity becomes $hQ$, which is of the order of unity in our simulations. \citet{Marzarietal2009} found that the disk self-gravity appears to be an important factor, because when included the circumprimary disk eccentricity evolution is considerably slower than in non self-gravitating disk models. We note, however, that \citet{Marzarietal2009} follow the disk eccentricity evolution only to $\sim0.012\,$Myr, which is too short to develop significant eccentricity in our models. In addition, \citet{Adamsetal1989} found that eccentric disk instabilities might also be excited by a process called as stimulation by the long-range interaction of Newtonian gravity for massive disks. Thus, the inclusion of self-gravity may also certainly improve our results.

For simplicity, we have assumed that the dust and gas temperature are well coupled in the disk atmosphere. However, \citet{Glassgoldetal2004} and \citet{KampDullemond2004} found that the dust and gas coupling ceases above the disk atmosphere. As a consequence, the gas in the tenuous region above the disk atmosphere might be significantly hotter than predicted by the double-layer model. Hence, the slowly rotating distant disk parcels above the disk atmosphere could produce a substantial contribution to the low-velocity part of the line profile, resulting in a centrally peaked profile. Nevertheless, the centrally peaked line profiles are expected to be asymmetric because of the supersonic Doppler-shift. To study the effect of the dust-gas temperature decoupling on the line profile asymmetry, a more sophisticated temperature model is needed.

We have neglected the irradiation flux coming from the secondary star. The luminosity of an $2.5\,\mathrm{Myr}$ old $0.3\,M_{\odot}$ mass secondary is $\sim 20\%$ that of the $1\,M_{\odot}$ primary \citep{Siessetal2000}. The final disk radius is $<0.45a_\mathrm{bin}$ on average (Fig.\,\ref{fig:ecc_r_multi}), while the disk eccentricity can reach $e\simeq0.4$ at the disk edge (Fig.\,\ref{fig:profile_evol}, panel d). Thus, the minimal distance between the disk apastron-edge and the secondary is $>0.37a_\mathrm{bin}$, taking into account the eccentric disk shape. This distance corresponds to 14.8\,AU and 7.5\,AU for model \#1 and model \#2, respectively. At these distances, the CO ro-vibrational fundamental band is not excited thermally by the secondary irradiation. Therefore, the effect of the secondary's irradiation on the CO line profiles should be investigated only for close-separation ($<20\,\mathrm{AU}$) binaries. 

A recent survey of Taurus medium-separation young binaries revealed that their secondary infrared fluxes are commensurable to their primary ones \citep{Pascuccietal2008}, hence, young binaries might have also circumsecondary disks. In some cases, the infrared flux of a circumsecondary disk is equal to that of the primary, and the common spectra of a system may also contain the presumably distorted line profiles of the secondary. To investigate these systems, more elaborate hydrodynamical simulations, incorporating the secondary's disk are needed.

As mentioned in Sect. 2., the secondary star induces a tidal wave during close encounters in each orbit. The gas is significantly compressed inside these temporary waves, as the density enhancement can reach two orders of magnitudes, according to our hydrodynamical simulations. Since the gas might be heated to temperatures high enough to excite the CO, the secondary mass flow might have a significant impact on the CO line profiles. Thus, to investigate the effect of the secondary mass flow on the line profile asymmetry, we need to incorporate the energy conservation in the 2D hydrodynamical simulations, which will be the subject of an upcoming paper.

\subsection{Outlook}

Spectroscopy of T\,Tauri stars detects the emission of molecules such as $\mathrm{H_2O}$, $\mathrm{OH}$, $\mathrm{HCN}$, $\mathrm{C_2H_2}$, and $\mathrm{CO_2}$. Nevertheless, CO is more abundant than any of these molecules by a factor of $\sim10$. Only $\mathrm{H_2O}$ could reach the abundance of CO predicted by recent models that calculate the vertical chemical structure of the gas in disk atmosphere (e.g., \citet{Glassgoldetal2004}, \citet{KampDullemond2004}, and \citet{Woitkeetal2009}). In some cases, such as AA\,Tau \citep{CarrNajita2008}, and both AS\,205A and DR\,Tau \citep{Salyketal2008}, the  rotational transitions of $\mathrm{H_2O}$ dominate the mid-infrared ($10-20\,\mathrm{\mu m}$) spectra, implying that $\mathrm{H_2O}$ is abundant in disk atmospheres. In contrast to theoretical predictions, strong water emission could be the consequence of turbulent mixing that carries molecules from disk midplane, where they are abundant, to the disk atmosphere \citep{CarrNajita2008}, or the effects of an enhanced mechanical heating of the atmosphere \citep{Glassgoldetal2009}. While the CO and $\mathrm{H_2O}$ ro-vibrational lines provide information about the inner regions of disk out to radii 2-3\,AU, their rotational lines are excited in the radii range 10-100\,AU \citep{Meijerinketal2008,BeckwithSargent1993}. As the disk eccentricity beyond 3\,AU might be large, $e_\mathrm{disk}\geq0.4$ (see, e.g., Fig.\,\ref{fig:profile_evol}, panel d), the rotational lines are also subject to large distortions owing to the disk eccentricity. Thus, it is worthwhile to search for asymmetric line profiles emerging from young binaries not only in the ro-vibrational spectra of CO and $\mathrm{H_2O}$, but also in the rotational spectra of CO and $\mathrm{H_2O}$. However, as the $\mathrm{H_2O}$ is heated by stellar X-rays and sub-thermally populated beyond ~0.3\,AU, X-ray heating and non-LTE level population treatment is needed to calculate water lines (e.g., \citet{Meijerinketal2008,Kampetal2010}). The calculation of asymmetric mid-infrared rotational molecular line profiles emerging from young binaries will be the subject of a future study. Another interesting disk diagnostic tool could be the [OI] $6300\,\AA$ line. The stellar UV photons incident on the disk atmosphere are thought to photodissociate OH and H2O, producing a non-thermal population of excited neutral oxygen that decays radiatively. As a result, [OI] emission lines emerge from the disk atmosphere, as reported by \citet{Ackeetal2005} for Herbig\,Ae/Be stars.

\section{Summary}

We have examined the effects of a secondary star on a gas-rich protoplanetary disk encircling the primary based on the previous work of \citet{Lubow1991,Lubow1991b} and \citet{Kleyetal2008}. While Kley at al. used spatially constant viscosity $\nu$, we have assumed a $\alpha$-type viscosity \citep{ShakuraSunyaev1973}, which has a spatial dependence given by $\nu\sim R^{1/2}$. According to our results, the circumprimary disk  eccentricity begins to increase after the first couple of hundreds of binary orbits. We have found that the disk eccentricity grows to a certain maximum value, begins to decline, and then stabilizes at $\bar{e}_\mathrm{disk}\simeq 0.2-0.35$. During the eccentricity growth, the disk extends and shrinks back to 0.35-0.45 times the binary separation. By the time the disk reaches a quasi-steady eccentric state, the surface density profile, the disk truncation radius, as well as the eccentricity profile no longer evolve. 

We have found that the final average disk eccentricity is independent of the binary mass ratio in the range $0.05\leq q\leq0.7$ (Fig.\,\ref{fig:ecc_r_multi}, panel a), of the magnitude of $\alpha$ in the range $0.005\leq\alpha\leq0.02$ (Fig.\,\ref{fig:ecc_r_multi}, panel b), of the disk aspect ratio for non-flaring models with $0.03\leq h\leq0.075$ (Fig.\,\ref{fig:ecc_r_multi}, panel d), of the disk-to-secondary mass ratio in the range $0.001\leq q_\mathrm{disk/sec}\leq0.1$ (Fig.\,\ref{fig:ecc_r_multi}, panel e), and of the choice of the inner boundary conditions (open outflow or rigid) (Fig.\,\ref{fig:ecc_r_multi}, panel f). In contrast, the disk eccentricity is lower for thick disks with aspect ratios above $h=0.1$, while disk eccentricity does not develop at all for thin disks with $h\leq0.01$ (Fig.\,\ref{fig:ecc_r_multi}, panel d). The disk average eccentricity is $\bar{e}_\mathrm{disk}\simeq 0.2$ for a flaring geometry with flaring index $\gamma=2/7$, while $\bar{e}_\mathrm{disk}\simeq 0.35$ for non-flaring models assuming a reasonable $h(a_\mathrm{bin})\simeq0.05$ for the disk aspect ratio (Fig.\,\ref{fig:aspect}). We emphasize that the disk becomes eccentric only temporarily for models assumed to have an eccentric binary orbit with $e_\mathrm{bin}\geq0.2$. Thus, the binary's orbital eccentricity protects the disk against eccentricity formation that might inhibit the planet formation.

In the eccentric disk, the orbit of gas parcels is non-circular, rather than elliptic. The average disk eccentricity is $\bar{e}_\mathrm{disk}\simeq0.2$ inside 2-3\,AU (see, Fig.\,\ref{fig:profile_evol}, panel d), where the ro-vibrational fundamental band of CO is excited by the primary's irradiation. Combining our hydrodynamical simulation with our semi-analytic double-layer disk model \citep{Regalyetal2010}, we have calculated the fundamental band $V=1\rightarrow0\,\mathrm{P10}$ ro-vibrational line profile of $\mathrm{^{12}C^{16}O}$ at $4.75\,\mathrm{\mu m}$ emerging from the superheated disk atmosphere. Since the emission spectra of the CO is strongly affected by the supersonic Doppler shift, the CO line profiles emerging from the eccentric disk are asymmetric departing from their well-known symmetric double-peaked form \citep{HorneMarsh1986}. We have found that the maximum peak-to-peak asymmetry is $A_\mathrm{pp}\simeq20\%$ for our models, which is above the detection limit of today's high-resolution mid-IR instruments such as CRIRES \citep{Kaeufletal2004}. The peak-to-peak asymmetry exhibits periodic variations as the disk precesses on the timescale of several times the binary period (Fig.\,\ref{fig:ts}). As the disk precession period is several hundred decades in medium-separation binaries, the detection of peak-to-peak asymmetry variations is improbable within a decade. On the other hand, the slight variations seen on the line wings (Fig.\,\ref{fig:ts}) have periods shorter than the binary period, which thus might be detected.

\section{Conclusion}

Our study of both the hydrodynamic and eccentricity evolution of the circumprimary disk in medium-separation young binary systems has revealed the following findings:

\begin{enumerate}
\item{The quasi-steady eccentric disk state always develops in circumprimary disks of young medium-separation ($20-40\,\mathrm{AU}$) binaries within the average disk lifetime, if the viscosity is between widely accepted values ($0.01\leq\alpha\leq0.1$).}
\item{The CO line profiles are asymmetric ($A_\mathrm{pp}\simeq20\%$) as the average disk eccentricity is $\bar{e}_\mathrm{disk}\simeq0.2$ inside 2-3\,AU, where the CO is excited by the primary's irradiation.}
\item{The orbital eccentricity of binary systems $e_\mathrm{bin}\geq0.2$ or their high/low disk geometrical thickness ($h\leq0.01$/$h\geq0.1$) might inhibit the development of the quasi-static disk eccentric state.}
\item{The inner ($R\leq2-3\,\mathrm{AU}$) disk eccentricity profile can be determined by fitting the observed high-resolution near-IR CO line profile asymmetry using a simple 2D spectral model.}
\end{enumerate}
\noindent 
Consequently, taking into account that the eccentricity of protoplanetary disks might strongly influence planet formation, by measuring it we might further constrain the planet formation theories in medium-separation binaries.

\begin{acknowledgements}
	This research has bee supported in part by DAAD-PPP mobility grant P-M\"OB/841/ and ``Lend\"ulet'' Young Researcher Program of the HAS. We would like to thank Bal\'azs Cs\'ak for his support in hydro-simulations.
\end{acknowledgements}


\bibliographystyle{aa}
\bibliography{regaly}

\end{document}